\documentclass[11pt,a4paper]{article}

\usepackage{amssymb}
\usepackage[dvips]{graphicx}
\usepackage{bm}

\begin{document}

\title{\bf Non-renormalization of the $V\bar cc$-vertices in ${\cal N}=1$ supersymmetric theories}

\author{
K.V.Stepanyantz\\
{\small{\em Moscow State University}}, {\small{\em  Physical
Faculty, Department  of Theoretical Physics}}\\
{\small{\em 119991, Moscow, Russia}}}

\maketitle

\begin{abstract}
Using the Slavnov--Taylor identities we prove that the three-point ghost vertices with a single line of the quantum gauge superfield are not renormalized in all loops in ${\cal N}=1$ supersymmetric gauge theories. This statement is verified by the explicit one-loop calculation made by the help of the BRST invariant version of the higher covariant derivative regularization. Using the restrictions to the renormalization constants which are imposed by the non-renormalization of the considered vertices we express the exact NSVZ $\beta$-function in terms of the anomalous dimensions of the Faddeev--Popov ghosts and of the quantum gauge superfield. In the expression for the NSVZ $\beta$-function obtained in this way the contributions of the Faddeev--Popov ghosts and of the matter superfields have the same structure.
\end{abstract}

\unitlength=1cm

keywords: supersymmetry, Slavnov--Taylor identities, renormalization, NSVZ $\beta$-function, higher covariant derivative regularization.

\section{Introduction}
\hspace*{\parindent}

Existence of ultraviolet divergences is a long standing problem of quantum field theory. Supersymmetry allows to improve the ultraviolet behaviour due to the so-call non-renormalization theorems. For example, ${\cal N}=4$ supersymmetric Yang--Mills (SYM) theory is finite in all orders \cite{Grisaru:1982zh,Mandelstam:1982cb,Brink:1982pd,Howe:1983sr}, and ${\cal N}=2$ supersymmetric theories are divergent only in the one-loop approximation \cite{Grisaru:1982zh,Howe:1983sr,Buchbinder:1997ib}. Using the ${\cal N}=2$ non-renormalization theorem it is possible to construct finite theories with ${\cal N}=2$ supersymmetry \cite{Howe:1983wj}. It is well known that the superpotential of ${\cal N}=1$ supersymmetric theories does not receive divergent quantum corrections \cite{Grisaru:1979wc}, and the $\beta$-function of these theories is related to the anomalous dimension by a special equation \cite{Novikov:1983uc,Jones:1983ip,Novikov:1985rd,Shifman:1986zi,Vainshtein:1986ja,Shifman:1985fi}, which is called the exact NSVZ $\beta$-function (or the NSVZ relation). For the ${\cal N}=1$ SYM theory without matter superfields the NSVZ equation gives the exact expression for the $\beta$-function, which appears to be a geometric progression.

The non-renormalization theorems appear due to large symmetries of a theory. Therefore, deriving them it is essential to assume that these symmetries remain unbroken at the quantum level. This means that one has to use an invariant regularization.\footnote{Non-invariant regularizations supplemented by a special subtraction scheme which restore the Slavnov--Taylor identities can be also used \cite{Slavnov:2001pu,Slavnov:2002ir,Slavnov:2002kg,Slavnov:2003cx}, but they are much more inconvenient.} In supersymmetric theories it is not a trivial problem \cite{Jack:1997sr}, because the dimensional regularization \cite{'tHooft:1972fi,Bollini:1972ui,Ashmore:1972uj,Cicuta:1972jf} breaks the supersymmetry \cite{Delbourgo:1974az}, while its modification called the dimensional reduction \cite{Siegel:1979wq} is not mathematically consistent \cite{Siegel:1980qs}. Removing of the inconsistencies leads to the loss of manifest supersymmetry \cite{Avdeev:1981vf} and to breaking supersymmetry by quantum corrections in higher loops \cite{Avdeev:1982xy,Avdeev:1982np,Velizhanin:2008rw}. Actually, the only invariant regularization which can keep supersymmetry and the gauge invariance unbroken is the higher covariant derivative regularization \cite{Slavnov:1971aw,Slavnov:1972sq}. In the supersymmetric case it can be formulated in the manifestly supersymmetric way in terms of ${\cal N}=1$ superfields \cite{Krivoshchekov:1978xg,West:1985jx}. It was also generalized to the case of ${\cal N}=2$ supersymmetry \cite{Krivoshchekov:1985pq,Buchbinder:2014wra}, but in order to have manifest ${\cal N}=2$ supersymmetry at all steps of quantum corrections calculating one should formulate the higher derivative regularization in ${\cal N}=2$ harmonic superspace \cite{Buchbinder:2001wy,Galperin:2001uw}. This was done in \cite{Buchbinder:2015eva} and allows to give a simple proof of the ${\cal N}=2$ non-renormalization theorem.

In this paper we investigate renormalization of theories with ${\cal N}=1$ supersymmetry, so that we will use the ${\cal N}=1$ supersymmetric BRST invariant version of the higher covariant derivative regularization. This regularization allows to calculate quantum corrections in a manifestly gauge and ${\cal N}=1$ supersymmetric way. An example of such a calculation can be found in \cite{Aleshin:2016yvj}, where the one-loop divergences have been obtained using this regularization. The result reveals an interesting feature of the quantum corrections: the three-point vertices with two ghost legs and one leg of the quantum gauge superfields are finite in the one-loop approximation. In this paper we prove that this fact is not accidental and follows from the Slavnov--Taylor identities \cite{Taylor:1971ff,Slavnov:1972fg} for the general renormalizable ${\cal N}=1$ supersymmetric gauge theory with matter. In principle, this statement can be considered as a new non-renormalization theorem in ${\cal N}=1$ supersymmetric theories. Moreover, it seems to be useful for deriving the exact NSVZ $\beta$-function by the direct summation of Feynman diagrams in the non-Abelian case.

\begin{figure}[h]
\begin{picture}(0,1.8)
\put(8.9,-2.5){\includegraphics[scale=0.4]{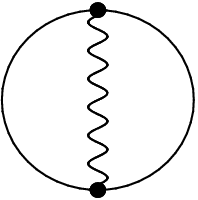}}
\put(7.1,-1.9){\includegraphics[scale=0.5]{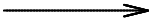}}
\put(10.7,-1.9){\includegraphics[scale=0.5]{arrow.eps}}
\put(6.5,-1.9){$\left.\vphantom{\begin{array}{c} 1\\ 1\\
1\\ 1\\ 1\\ 1\\ 1\\ 1\\ 1\\ 1\\ 1\\
\end{array}} \right\}$}
\put(0,0.2){\includegraphics[scale=0.15]{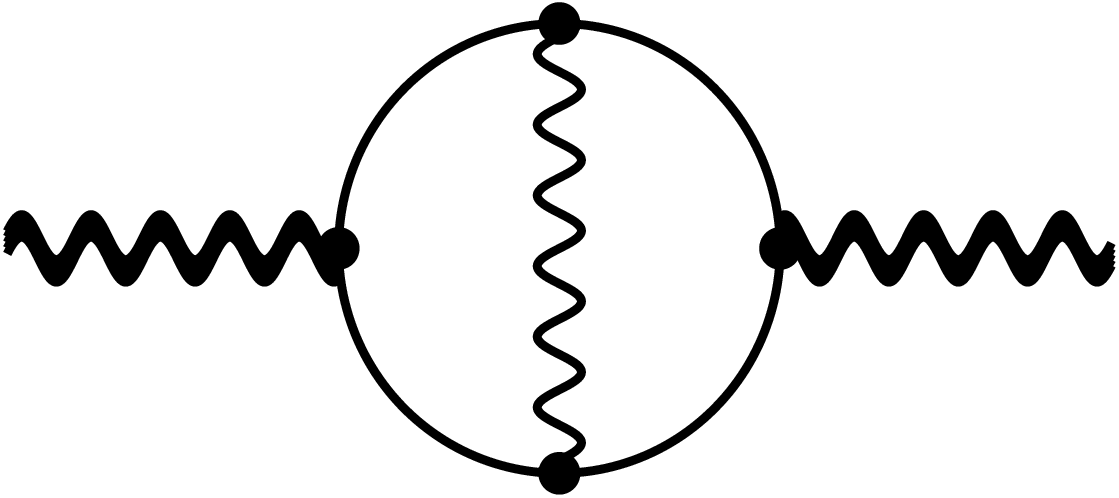}}
\put(3.4,0.2){\includegraphics[scale=0.15]{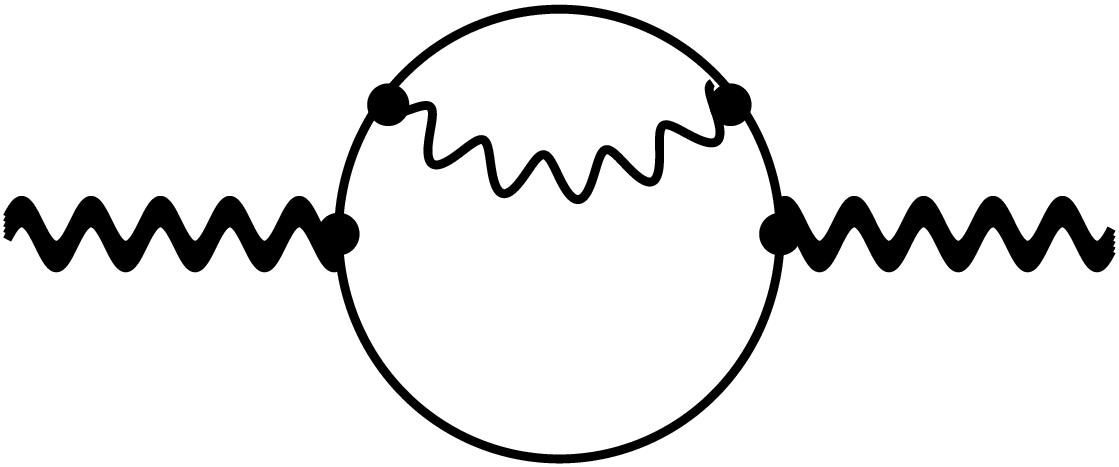}}
\put(0,-2.05){\includegraphics[scale=0.15]{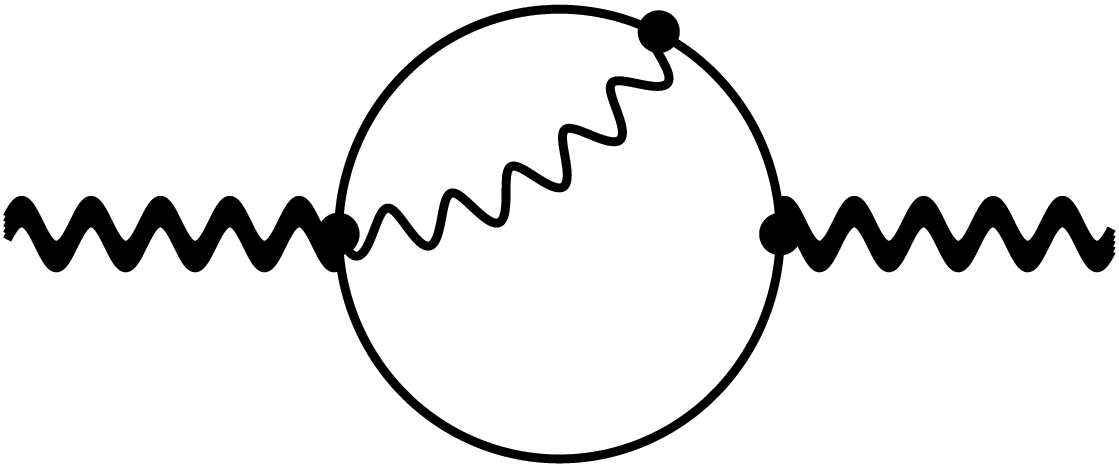}}
\put(3.4,-2.05){\includegraphics[scale=0.15]{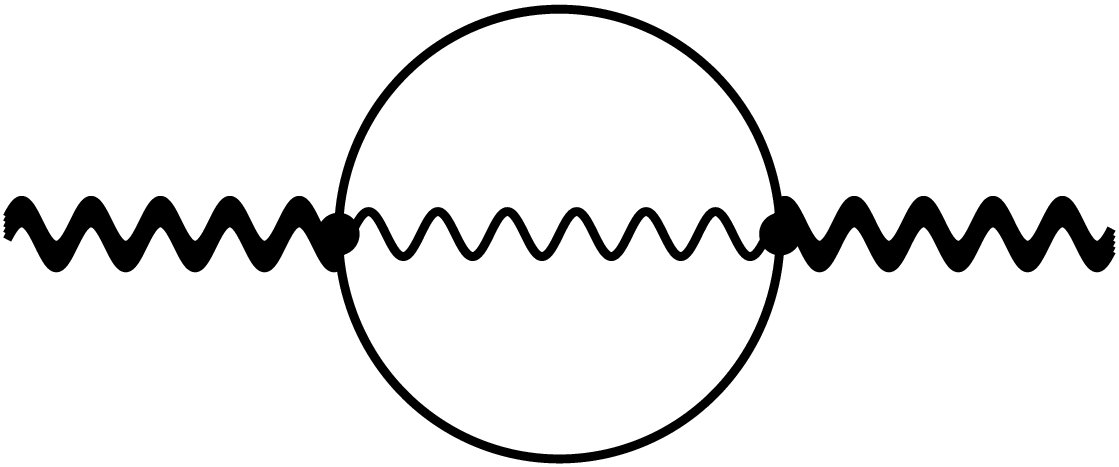}}
\put(0.27,-4.6){\includegraphics[scale=0.148]{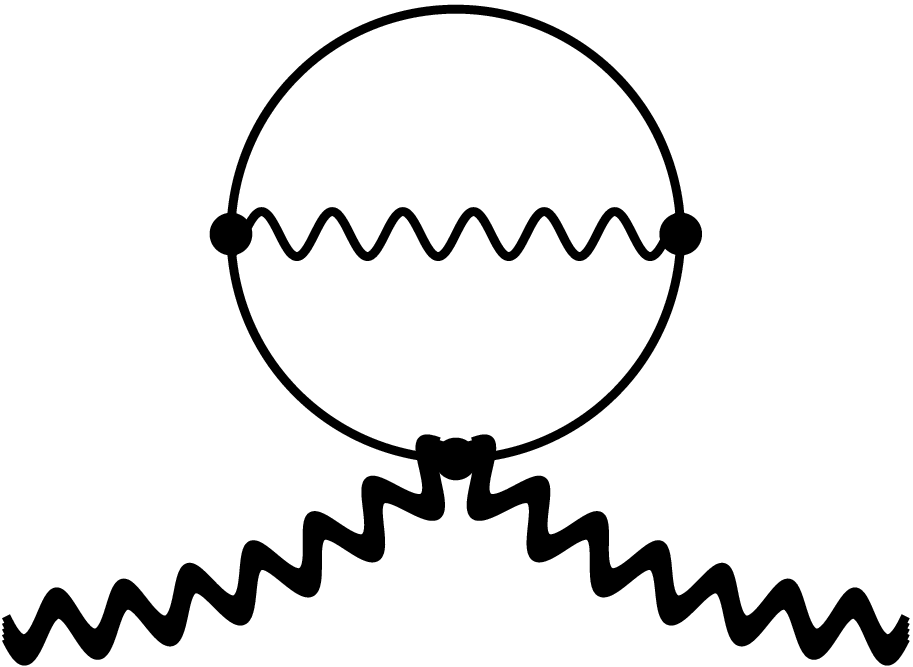}}
\put(3.7,-4.6){\includegraphics[scale=0.148]{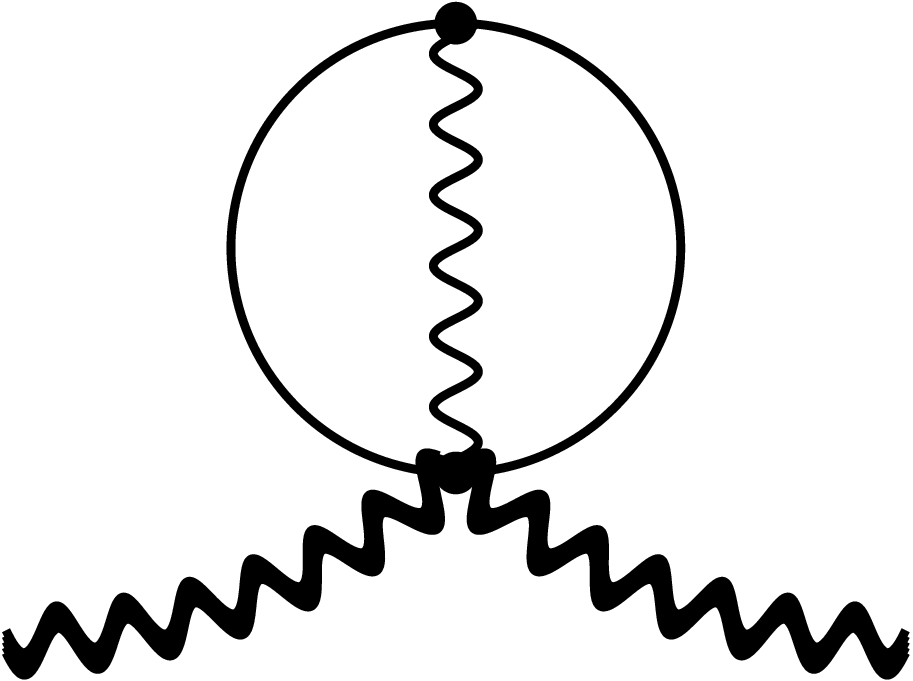}}
\put(12.5,-2.5){\includegraphics[scale=0.4]{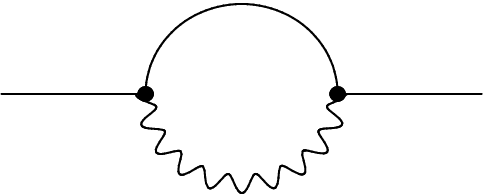}}
\end{picture}
\vspace*{4.8cm} \caption{This figure qualitatively illustrates how the diagrams contributing to the $\beta$-function produce the corresponding diagrams contributing to the anomalous dimension. (The bold external lines corresponds to the background gauge superfield $\bm{V}$, while the thin wavy lines correspond to the quantum gauge superfield. The matter propagators are denoted by the usual line.)}\label{Figure_Qualitative_Picture}
\end{figure}

In the Abelian case the NSVZ relation was obtained by the direct summation of Feynman diagrams in all orders for the renormalization group (RG) functions defined in terms of the bare coupling constant in \cite{Stepanyantz:2011jy,Stepanyantz:2014ima}. A similar expression for the Adler $D$-function \cite{Adler:1974gd} in ${\cal N}=1$ SQCD was also derived in \cite{Shifman:2014cya,Shifman:2015doa}. Both these derivations are based on the observation that the integrals giving the $\beta$-function (defined in terms of the bare coupling constant) in supersymmetric theories are integrals of (double) total derivatives in the momentum space \cite{Soloshenko:2003nc,Smilga:2004zr}. This structure of loop integrals was confirmed by a large number of explicit loop calculations  (see, e.g. \cite{Pimenov:2009hv,Stepanyantz:2011bz,Stepanyantz:2012zz,Stepanyantz:2012us,Kazantsev:2014yna,Buchbinder:2014wra,Buchbinder:2015eva}). It allows calculating one of the loop integrals analytically and relating renormalization of the coupling constant in a certain order to the renormalization of the matter superfields in the previous order. Qualitatively this picture is illustrated by Fig. \ref{Figure_Qualitative_Picture} \cite{Smilga:2004zr,Kazantsev:2014yna,Pimenov:2006cu}. From the left we present two-loop diagrams contributing to the $\beta$-function. They contain two external lines of the background gauge superfield attached to the same two-loop vacuum graph, which is shown in the center of the figure. Cutting the matter line in this graph we obtain the one-loop diagram contributing to the anomalous dimension of the matter superfields. The detailed discussion of the corresponding results in the three-loop approximation can be found in \cite{Kazantsev:2014yna}.

In the non-Abelian case cuts of internal lines will give the diagrams contributing to the anomalous dimensions of the matter superfields, of the quantum gauge superfield, and of the Faddeev--Popov ghosts. Therefore, it is desirable to express the $\beta$-function in terms of these anomalous dimensions. In this paper we will demonstrate that this can be done using the non-renormalization theorem for the three-point vertices with two ghost lines and one line of the quantum gauge superfield. In particular, we will see that chiral ghosts superfields and chiral matter superfields similarly contribute to the NSVZ $\beta$-function. Thus, the statement derived in this paper may occur very useful for deriving the NSVZ relation in the non-Abelian case.

This paper is organized as follows: In Sect. \ref{Section_STI} we consider a general renormalizable ${\cal N}=1$ SYM theory with matter. We regularize it by the BRST invariant version of the higher covariant derivative regularization in order that supersymmetry, the background gauge symmetry, and the BRST symmetry will be unbroken at the quantum level. Then the Slavnov--Taylor identities are constructed using the BRST invariance of the (regularized) theory. In Sect. \ref{Section_3Point_STI} these Slavnov--Taylor identities are written for the three-point ghost-gauge vertices. Using them in Sect. \ref{Section_Non_Renormalization} we prove the finiteness of these vertices. (Note that we consider the vertices with the quantum gauge superfield.) This statement is verifies by the explicit one-loop calculation in Sect. \ref{Section_One_Loop}. Finally, in Sect. \ref{Section_NSVZ} using the non-renormalization theorem proved in this paper we rewrite the NSVZ relation for the ${\cal N}=1$ non-Abelian SYM theories in such a form that the Faddeev--Popov ghosts, the quantum gauge superfield, and the chiral matter superfields similarly contribute to the right hand side. In this section we also discuss, why this form of the NSVZ relation is useful for making a general prove of the exact NSVZ $\beta$-function by summing the Feynman diagrams. Also here we suggest the conditions that define the NSVZ scheme in the non-Abelian case with the BRST-invariant version of the higher covariant derivative regularization, if the RG functions are defined in terms of the renormalized coupling constants.

\section{${\cal N}=1$ supersymmetric gauge theories and the Slavnov--Taylor identities}
\hspace*{\parindent}\label{Section_STI}

In this paper we consider the general renormalizable ${\cal N}=1$ SYM theory, which is described by the action

\begin{eqnarray}\label{Action_Classical}
&& S = \frac{1}{2 e_0^2}\,\mbox{Re}\,\mbox{tr}\int d^4x\,
d^2\theta\,W^a W_a + \frac{1}{4} \int d^4x\, d^4\theta\,\phi^{*i}
(e^{2V})_i{}^j \phi_j\nonumber\\
&&\qquad\qquad\qquad\qquad\qquad\qquad  +
\Bigg\{\int d^4x\,d^2\theta\,\Big(\frac{1}{4} m_0^{ij} \phi_i \phi_j + \frac{1}{6}\lambda_0^{ijk} \phi_i
\phi_j \phi_k\Big) + \mbox{c.c.}\Bigg\},\qquad
\end{eqnarray}

\noindent
where $e_0$, $m_0^{ij}$, and $\lambda_0^{ijk}$ are the bare coupling constant, the mass matrix, and the Yukawa constants, respectively. $V=e_0 V^A T^A$ is the gauge superfield and $\phi_i$ are chiral matter superfields in a certain representation $R$ of the gauge group. The chiral superfield

\begin{equation}
W_a = \frac{1}{8} \bar D^2 \left(e^{-2V} D_a e^{2V}\right)
\end{equation}

\noindent
is the supersymmetric gauge field strength, where $D_a$ is the (right) supersymmetric covariant derivative. The left supersymmetric covariant derivative is denoted by $\bar D_{\dot a}$. In our notation, the generators of the fundamental representation $t^A$ are normalized by the equation $\mbox{tr}(t^A t^B) = \delta^{AB}/2$. The generators of the representation $R$ we denote by $T^A$. Under the assumption that the masses and Yukawa constants satisfy the equations

\begin{eqnarray}
&& m_0^{ik} (T^A)_k{}^{j} + m_0^{kj} (T^A)_k{}^{i} = 0;\vphantom{\Big(}\nonumber\\
&& \lambda_0^{ijm} (T^A)_m{}^{k} + \lambda_0^{imk} (T^A)_m{}^{j} +
\lambda_0^{mjk} (T^A)_m{}^{i} = 0,\vphantom{\Big(}
\end{eqnarray}

\noindent
the considered theory is invariant under the gauge transformations

\begin{equation}\label{Gauge_Transformations}
\phi \to e^{A}\phi;\qquad e^{2V} \to e^{-A^+} e^{2V} e^{-A},
\end{equation}

\noindent
where $A = i e_0 A^B T^B$ is an arbitrary chiral superfield in the adjoint representation of the gauge group. We also define the superfield $\Omega$ as a solution of the equation $e^{2V} \equiv e^{\Omega^+} e^{\Omega}$. Using this superfield one can introduce the background field method \cite{DeWitt:1965jb,Abbott:1980hw,Abbott:1981ke} by making the substitution

\begin{equation}
e^{\Omega} \to e^{\Omega} e^{\bm{\Omega}}.
\end{equation}

\noindent The background gauge superfield $\bm{V}$ is then defined by the equation
$e^{2\bm{V}} = e^{\bm{\Omega}^+} e^{\bm{\Omega}}$, and $V$ becomes a quantum gauge superfield.
The background field method enables us to construct the effective action manifestly invariant under the background gauge transformations

\begin{equation}\label{Background_Gauge_Invariance_Transformations}
e^{\bm{\Omega}} \to e^{iK} e^{\bm{\Omega}} e^{-A};\qquad e^{\Omega}
\to e^{\Omega} e^{-iK};\qquad V\to e^{iK} V e^{-iK};\qquad \phi \to
e^A \phi,
\end{equation}

\noindent where the parameter $K$ is a hermitian superfield. The quantum gauge invariance is broken by the gauge fixing procedure, the remaining symmetry being the BRST invariance \cite{Becchi:1974md,Tyutin:1975qk}.

At the quantum level the gauge invariance is encoded in the Slavnov--Taylor identities \cite{Taylor:1971ff,Slavnov:1972fg}. These identities follows from the BRST invariance of the full action which also includes a gauge fixing term and ghosts. That is why we will regularize the considered theory in such a way that the BRST invariance is unbroken. Certainly, it is also highly desirable that supersymmetry is also unbroken. Both these requirements can be satisfied if the higher covariant derivative method is used for regularization. Following Ref. \cite{Aleshin:2016yvj}, we add to the classical action (\ref{Action_Classical}) the term containing the higher covariant derivatives

\begin{eqnarray}\label{Action_HD_Term}
&& S_{\Lambda} = \frac{1}{2e_0^2}\,\mbox{Re}\,\mbox{tr}\int d^4x\,
d^2\theta\, e^{\Omega} e^{\bm{\Omega}} W^a e^{-\bm{\Omega}}
e^{-\Omega} \Big[R\Big(-\frac{\bar\nabla^2
\nabla^2}{16\Lambda^2}\Big) -1\Big]_{Adj} e^{\Omega} e^{\bm{\Omega}}
W_a
e^{-\bm{\Omega}} e^{-\Omega}\qquad\nonumber\\
&& + \frac{1}{4} \int d^4x\, d^4\theta\,\phi^+ e^{\bm{\Omega}^+}
e^{\Omega^+} \Big[F\Big(-\frac{\bar\nabla^2
\nabla^2}{16\Lambda^2}\Big)-1\Big] e^{\Omega} e^{\bm{\Omega}}
\phi,
\end{eqnarray}

\noindent
where the regulators $R$ and $F$ rapidly grow at the infinity. Consequently, the propagators contain large degrees of momentums in the denominator and all diagrams beyond the one-loop approximation \cite{Faddeev:1980be} become convergent (except for the one-loop subdivergencies). The remaining one-loop divergencies and subdivergencies should be regularized by inserting the Pauli--Villars determinants into the generating functional \cite{Slavnov:1977zf}. Then the generating functional can be written as

\begin{eqnarray}
&& Z[\bm{V},\mbox{Sources}] = \int D\mu\,
\mbox{Det}(PV,M_{\Phi}) \mbox{Det}(PV,M_{\varphi})^{-1}
\qquad\nonumber\\
&&\qquad\qquad\qquad\qquad\qquad \times \exp\Big(iS + iS_\Lambda +
iS_{\mbox{\scriptsize gf}} + i S_{\mbox{\scriptsize FP}} + i
S_{\mbox{\scriptsize NK}} + i S_{\mbox{\scriptsize
sources}}\Big),\qquad
\end{eqnarray}

\noindent where $D\mu$ denotes the measure of the functional integration. The gauge fixing term has the form

\begin{eqnarray}\label{Gauge_Fixing_With_F}
&& S_{\mbox{\scriptsize gf}} = \frac{1}{e_0^2} \mbox{tr}\int
d^4x\,d^4\theta\,\Big(16 \xi_0\,  f^+ \Big[e^{\bm{\Omega^+}}
K^{-1}\Big(-\frac{\bm{\bar\nabla}^2 \bm{\nabla}^2}{16\Lambda^2}\Big)
e^{\bm{\Omega}}\Big]_{Adj} f \nonumber\\
&& \qquad\qquad\qquad\qquad\qquad\qquad\qquad\qquad +
e^{\bm{\Omega}} f e^{-\bm{\Omega}} \bm{\nabla}^2 V +
e^{-\bm{\Omega^+}} f^+ e^{\bm{\Omega^+}} \bm{\bar \nabla}^2 V
\Big),\qquad
\end{eqnarray}

\noindent
where $f$ is a commuting chiral superfield in the adjoint representation of the gauge group, and is evidently invariant under the background gauge transformations (\ref{Background_Gauge_Invariance_Transformations}). The corresponding actions for the Faddeev--Popov and Nielsen--Kallosh ghosts are given by the expressions

\begin{eqnarray}
&& S_{\mbox{\scriptsize FP}} = \frac{1}{e_0^2} \mbox{tr} \int
d^4x\,d^4\theta\, \left(e^{\bm{\Omega}}\bar c e^{-\bm{\Omega}} +
e^{-\bm{\Omega}^+}\bar c^+ e^{\bm{\Omega}^+}\right)\nonumber\\
&&\qquad\qquad\qquad\quad \times \Big\{
\Big(\frac{V}{1-e^{2V}}\Big)_{Adj} \Big(e^{-\bm{\Omega}^+} c^+
e^{\bm{\Omega}^+}\Big) + \Big(\frac{V}{1-e^{-2V}}\Big)_{Adj}
\Big(e^{\bm{\Omega}} c
e^{-\bm{\Omega}}\Big)\Big\};\qquad\\
&& S_{\mbox{\scriptsize NK}} = \frac{1}{2e_0^2}\mbox{tr}\int
d^4x\,d^4\theta\,b^+ \Big[e^{\bm{\Omega}^+}
K\Big(-\frac{\bm{\bar\nabla}^2 \bm{\nabla}^2}{16\Lambda^2}\Big)
e^{\bm{\Omega}}\Big]_{Adj} b,
\end{eqnarray}

\noindent
respectively. They are also invariant under the transformations (\ref{Background_Gauge_Invariance_Transformations}). The sources can be written as

\begin{eqnarray}
&& S_{\mbox{\scriptsize Sources}} = \int d^4x\, d^2\theta\, \Big(j_c^A c^A + \bar j_c^A \bar c^A + j^i \phi_i\Big)
\nonumber\\
&& \qquad\qquad\qquad
+ \int d^4x\, d^2\bar\theta\, \Big(c^{*A} j_c^{*A} + \bar c^{*A} \bar j_c^{*A} + j_i^* \phi^{*i}\Big) + \int d^4x\,d^4\theta\,V^A J^A.\qquad
\end{eqnarray}

The total action of the gauge fixed theory is invariant under the BRST transformations \cite{Becchi:1974md,Tyutin:1975qk} which in the supersymmetric case have the form

\begin{eqnarray}\label{BRST_Transformation}
&& \delta V = - \varepsilon  \Big\{
\Big(\frac{V}{1-e^{2V}}\Big)_{Adj} \left(e^{-\bm{\Omega}^+} c^+
e^{\bm{\Omega}^+}\right)  + \Big(\frac{V}{1-e^{-2V}}\Big)_{Adj}
\left(e^{\bm{\Omega}} c e^{-\bm{\Omega}}\right)\Big\};\qquad
\delta \phi = \varepsilon c \phi;\nonumber\\
&& \delta \bar c = \varepsilon \bar D^2 (e^{-2\bm{V}} f^+
e^{2\bm{V}});\qquad\, \delta \bar c^+ = \varepsilon D^2(e^{2\bm{V}}f
e^{-2\bm{V}});\qquad \delta c = \varepsilon c^2;\qquad \delta c^+ =
\varepsilon (c^+)^2;
\vphantom{\frac{\Lambda^2}{\Lambda^2}}\quad\nonumber\\
&&\qquad\qquad\quad \delta f = \delta f^+ = 0; \qquad\quad \delta b
= \delta b^+ = 0; \qquad\quad \delta \bm{\Omega} =
\delta\bm{\Omega}^+ = 0,\vphantom{\frac{\Lambda^2}{\Lambda^2}}
\end{eqnarray}

The Slavnov--Taylor identity can be obtained by making the substitution (\ref{BRST_Transformation}) in the generating functional and can be written in the form

\begin{eqnarray}\label{Slavnov_Taylor_Identity}
&& \int d^4x\, d^4\theta_x\, \frac{\delta\Gamma}{\delta V_x^A}\left\langle\delta V_x^A \right\rangle
+ \int d^4x\, d^2\theta_x\, \Big(\left\langle\delta \bar c_x^A \right\rangle\frac{\delta\Gamma}{\delta \bar c_x^A}
+ \left\langle\delta c_x^A \right\rangle\frac{\delta\Gamma}{\delta c_x^A} + \left\langle\delta \phi_i \right\rangle\frac{\delta\Gamma}{\delta \phi_i} \Big)
\nonumber\\
&& + \int d^4x\, d^2\bar\theta_x\, \Big(\left\langle\delta \bar c_x^{*A} \right\rangle\frac{\delta\Gamma}{\delta \bar c_x^{*A}}
+ \left\langle\delta c_x^{*A} \right\rangle\frac{\delta\Gamma}{\delta c_x^{*A}} + \left\langle\delta \phi^{*i} \right\rangle\frac{\delta\Gamma}{\delta \phi^{*i}} \Big) = 0,\qquad
\end{eqnarray}

\noindent
where, for simplicity, we keep the dependence on $\varepsilon$. Note that here we use the notation

\begin{eqnarray}
&& \langle X(\mbox{fields}) \rangle \equiv \frac{1}{Z} \int D\mu\, X(\mbox{fields})\,
\mbox{Det}(PV,M_{\Phi}) \mbox{Det}(PV,M_{\varphi})^{-1}
\qquad\nonumber\\
&&\qquad\qquad\qquad\qquad\qquad \times \exp\Big(iS + iS_\Lambda +
iS_{\mbox{\scriptsize gf}} + i S_{\mbox{\scriptsize FP}} + i
S_{\mbox{\scriptsize NK}} + i S_{\mbox{\scriptsize
sources}}\Big),\qquad
\end{eqnarray}

\noindent
where the sources should be expressed in terms of fields in the standard way.

In this paper we are interested in diagrams which do not contain external lines of the background superfield. Therefore, below we set the background field to 0.
In this case after eliminating the auxiliary superfields $f$ and $f^*$ we obtain

\begin{equation}\label{Delta_Bar_C}
\left\langle \delta\bar c_x^A \right\rangle = -\frac{1}{16\xi_0}\varepsilon K\left(\partial^2/\Lambda^2\right) \bar D^2 D^2 V^A
\end{equation}

Also we will use one more identity which can be derived by making the substitution $\bar c \to \bar c + a$, where $a$ is an arbitrary chiral superfield. After this substitution and differentiating the result with respect to $a$ we obtain the first of the following identities:

\begin{eqnarray}\label{Ghost_Identities}
\varepsilon \frac{\delta\Gamma}{\delta\bar c_x^A} = \frac{1}{4}\bar D^2 \left\langle\delta V_x^A\right\rangle;\qquad
\varepsilon \frac{\delta\Gamma}{\delta\bar c_x^{*A}} = \frac{1}{4} D^2 \left\langle\delta V_x^A\right\rangle,
\end{eqnarray}

\noindent
where the background gauge superfield is also set to 0. The second identity can be found by the similar method, if one make the substitution $\bar c^+ \to \bar c^+ + a^+$. (These identities are well-known. For the (non-supersymmetric) Yang--Mills theory they are derived, e.g., in \cite{Itzykson:1980rh}.)

The Slavnov--Taylor identities allow proving the renormalizability of the supersymmetric gauge theories \cite{Slavnov:1974dg,Ferrara:1975ye,Piguet:1975md,Piguet:1981hh}. As a consequence, all divergencies can be absorbed into the renormalization of superfields and coupling constants. In our notation, the renormalization constants are defined by the following equations

\begin{eqnarray}\label{Z_Definition}
&& \frac{1}{\alpha_0} = \frac{Z_\alpha}{\alpha};\qquad
\frac{1}{\xi_0} = \frac{Z_\xi}{\xi};\qquad \bm{V} =
\bm{V}_R;\qquad V = Z_V Z_\alpha^{-1/2} V_R;\qquad b = \sqrt{Z_b} b_R;\qquad\nonumber\\
&& \bar c c = Z_c Z_\alpha^{-1} \bar c_R c_R;\qquad\ \ \ \phi_i = (\sqrt{Z_\phi})_i{}^j (\phi_R)_j;\qquad \ \
m^{ij} = m_0^{mn} (Z_m)_m{}^i (Z_m)_n{}^j;
\qquad\vphantom{\frac{Z_\xi}{\xi}}\\
&&\qquad\qquad\qquad\qquad\qquad \lambda^{ijk} = \lambda_0^{mnp} (Z_\lambda)_m{}^i (Z_\lambda)_n{}^j
(Z_\lambda)_p{}^k,\vphantom{\frac{Z_\xi}{xi}}\nonumber
\end{eqnarray}

\noindent
similarly to Ref. \cite{Aleshin:2016yvj}. Here the subscript $R$ denotes renormalized superfields, $\alpha$, $\lambda$, and $\xi$ are the renormalized coupling constant, Yukawa constant, and gauge parameter, respectively; $m$ denotes the renormalized masses. We can impose the following conditions on these renormalization constants:

\begin{equation}\label{Z_Constants_Natural}
(Z_m)_i{}^j = (Z_\lambda)_i{}^j = (\sqrt{Z_\phi})_i{}^j;\qquad Z_\xi = Z_V^{-2};\qquad Z_b = Z_\alpha^{-1}.
\end{equation}

\noindent
It is possible due to the non-renormalization of the superpotential, transversality of quantum corrections to the two-point Green function of the (quantum) gauge superfield, and structure of the Nielsen--Kallosh ghost Lagrangian, respectively. In general, these conditions are not mandatory, because finite renormalizations are also possible. However, these relation are natural and convenient. That is why below we will assume that they are always valid.

\section{Slavnov--Taylor identities for the $V\bar c c$ vertices}
\hspace*{\parindent}\label{Section_3Point_STI}

In this section we obtain the Slavnov--Taylor identities for the three-point vertices of the $V \bar c c$-type, which are a key ingredient for proving the non-renormalization theorem for these vertices.

Let us differentiate the Slavnov--Taylor identity (\ref{Slavnov_Taylor_Identity}) with respect to $\bar c_y^{*B}$, $c_z^C$, and $c_w^D$. After this, we set all fields equal to 0 and take into account that, due to the symmetry leading to the ghost number conservation, only Green functions with equal numbers of ghost and antighost legs can be non-trivial. The result has the form

\begin{eqnarray}\label{STI_Consequence}
&& 0 = \int d^4x\, d^4\theta\, \Big(\frac{\delta^3\Gamma}{\delta \bar c_y^{*B} \delta V_x^A \delta c_z^C} \cdot
\frac{\delta}{\delta c_w^D}\left\langle\delta V_x^A \right\rangle - \frac{\delta^3\Gamma}{\delta \bar c_y^{*B} \delta V_x^A \delta c_w^D} \cdot
\frac{\delta}{\delta c_z^C}\left\langle\delta V_x^A \right\rangle \Big)\qquad\nonumber\\
&& - \int d^4x\, d^2\theta\, \frac{\delta^2\Gamma}{\delta \bar c_y^{*B} \delta c_x^A}\cdot \frac{\delta^2}{\delta c_z^C \delta c_w^D}\left\langle \delta c_x^A\right\rangle,\qquad
\end{eqnarray}

\noindent
where we take into account that the ghost superfields are anticommuting. The derivatives of $\left\langle\delta V^A \right\rangle$ entering this equation can be expressed via the two-point Green functions of the Faddeev--Popov ghosts, which, due to the (anti)chirality of the ghost and antighost can be written in the form

\begin{equation}\label{G_C_Definition}
\frac{\delta^2\Gamma}{\delta \bar c_y^{*B} \delta c_x^A} = - \frac{D_y^2 \bar D_x^2}{16} G_c(\partial^2/\Lambda^2) \delta^8_{xy} \delta_{AB}; \qquad \quad \frac{\delta^2\Gamma}{\delta \bar c_y^{B} \delta c_x^{*A}} = \frac{\bar D_y^2 D_x^2}{16} G_c(\partial^2/\Lambda^2) \delta^8_{xy} \delta_{AB}.
\end{equation}

\noindent
(Note that the dimensionless function $G_c(\partial^2/\Lambda^2)$ is normalized in such a way that in the tree approximation $G_c=1$.) Really, from dimensional considerations and using chirality of the ghost superfields, we see that the expression $\delta \left\langle\delta V_y^B \right\rangle/\delta c_x^A$ is proportional to $\bar D_x^2\delta^8_{xy}$. Then taking into account the identities (\ref{Ghost_Identities}) we obtain

\begin{equation}\label{DeltaV_Derivative}
\frac{\delta}{\delta c_x^A}\left\langle\delta V_y^B \right\rangle = -\frac{\bar D_y^2 D_y^2}{16\partial^2} \frac{\delta}{\delta c_x^A}\left\langle\delta V_y^B \right\rangle = -\varepsilon\cdot \frac{\bar D_y^2}{4\partial^2} \frac{\delta^2\Gamma}{\delta \bar c_y^{*B} \delta c_x^A} = -\varepsilon\cdot \frac{1}{4} G_c(\partial^2/\Lambda^2)\,\bar D^2\delta^8_{xy} \delta_{AB}.
\end{equation}

\noindent
Similarly,

\begin{equation}
\frac{\delta}{\delta c_x^{*A}}\left\langle\delta V_y^B \right\rangle = \varepsilon\cdot \frac{1}{4} G_c(\partial^2/\Lambda^2)\, D^2\delta^8_{xy} \delta_{AB}.
\end{equation}

\noindent
Substituting the expressions (\ref{G_C_Definition}) and (\ref{DeltaV_Derivative}) into the Slavnov--Taylor identity (\ref{STI_Consequence}) we derive the following identity relating the three-point Green functions:

\begin{eqnarray}\label{STI_Ghosts1}
&&\varepsilon\cdot G_c(\partial_w^2/\Lambda^2) \bar D_w^2 \frac{\delta^3\Gamma}{\delta \bar c_y^{*B} \delta V_w^D \delta c_z^C} - \varepsilon \cdot G_c(\partial_z^2/\Lambda^2) \bar D_z^2 \frac{\delta^3\Gamma}{\delta \bar c_y^{*B} \delta V_z^C \delta c_w^D} \nonumber\\
&&\qquad\qquad\qquad\qquad\qquad\qquad\qquad\qquad + \frac{1}{2} G_c\left(\partial_y^2/\Lambda^2\right) D_y^2 \frac{\delta^2}{\delta c_z^C \delta c_w^D}\left\langle \delta c_y^B\right\rangle = 0.\qquad
\end{eqnarray}

Similarly differentiating the Slavnov--Taylor identity (\ref{Slavnov_Taylor_Identity}) with respect to $\bar c_y^{*B}$, $c_z^{*C}$, and $c_w^D$ we obtain

\begin{eqnarray}\label{STI_Ghosts2}
&&\varepsilon\cdot G_c(\partial_w^2/\Lambda^2) \bar D_w^2 \frac{\delta^3\Gamma}{\delta \bar c_y^{*B} \delta V_w^D \delta c_z^{*C}} + \varepsilon \cdot G_c(\partial_z^2/\Lambda^2) D_z^2 \frac{\delta^3\Gamma}{\delta \bar c_y^{*B} \delta V_z^C \delta c_w^D} \nonumber\\
&&\qquad\qquad\qquad\qquad\qquad\qquad\qquad\qquad + \frac{1}{2} G_c\left(\partial_y^2/\Lambda^2\right) D_y^2 \frac{\delta^2}{\delta c_z^{*C} \delta c_w^D}\left\langle \delta c_y^B\right\rangle = 0.\qquad
\end{eqnarray}

In order to simplify these identities we will use explicit expressions for the Green functions entering Eqs. (\ref{STI_Ghosts1}) and (\ref{STI_Ghosts2}). They can be obtained using dimensional and chirality considerations. It is convenient to present the result in the momentum representation using the notation

\begin{equation}
\delta^8_{xy}(p) \equiv \delta^4(\theta_x-\theta_y) e^{ip_\alpha (x^\alpha - y^\alpha)}.
\end{equation}

\noindent
Then the considered three-point ghost-gauge Green functions can be written in the form

\begin{eqnarray}\label{Three_Point_Function1}
&&\hspace*{-5mm} \frac{\delta^3\Gamma}{\delta \bar c_x^{*A} \delta V_y^B \delta c_z^C} = -\frac{i e_0}{16} f^{ABC} \int \frac{d^4p}{(2\pi)^4} \frac{d^4q}{(2\pi)^4} \Big(f(p,q) \partial^2\Pi_{1/2} - F_\mu(p,q) (\gamma^\mu)_{\dot a}{}^{b} \bar D^{\dot a} D_b\nonumber\\
&&\hspace*{-5mm}  + F(p,q) \Big)_{y} \Big(D_{x}^2\delta^8_{xy}(q+p)\, \bar D_{z}^2 \delta^8_{yz}(q)\Big);\\
&&\vphantom{1} \nonumber\\
\label{Three_Point_Function2}
&&\hspace*{-5mm}  \frac{\delta^3\Gamma}{\delta \bar c_x^{*A} \delta V_y^B \delta c_z^{*C}} = -\frac{i e_0}{16} f^{ABC} \int \frac{d^4p}{(2\pi)^4} \frac{d^4q}{(2\pi)^4} \widetilde F(p,q) D_x^2\delta^8_{xy}(q+p) D_z^2 \delta^8_{yz}(q),
\end{eqnarray}

\noindent
where $\partial^2 \Pi_{1/2}\equiv -D^a \bar D^2 D_a/8$ is the supersymmetric transversal projection operator. The functions $F(p,q)$, $\widetilde F(p,q)$, $F_\mu(p,q)$, and $f(p,q)$ can be found by calculating the corresponding Feynman diagrams. Explicit expressions for them in the one-loop approximation are presented below. Two remaining Green functions of the type $V\bar c c$ (which were not written above) are obtained by the complex conjugation and are expressed in terms of $F$, $\widetilde F$, $F_\mu$, and $f$ in the same way.

Also we need two correlators containing derivatives of $\left\langle \delta c^B\right\rangle = \varepsilon\cdot ie_0 f^{BCD}\left\langle c^C c^D\right\rangle/2$, which enter into Eqs. (\ref{STI_Ghosts1}) and (\ref{STI_Ghosts2}). In order to investigate them let us introduce the (Grassmannian even) chiral source superfield ${\cal J}$ for the product of ghost superfields by adding the term

\begin{equation}\label{Ghost_Source}
-\frac{e_0}{2} \int d^4x\, d^2\theta\, f^{ABC} {\cal J}^A c^B c^C +\mbox{c.c.}
\end{equation}

\noindent
to the classical action. Due to the nilpotency of the BRST transformations this term evidently does not break the BRST invariance. Then it is easy to see that the considered Green functions can be presented as

\begin{equation}\label{Ghost_Green_Functions}
\frac{\delta^2}{\delta c_z^C \delta c_w^D}\left\langle \delta c_y^B\right\rangle = -i \varepsilon\cdot \frac{\delta^3\Gamma}{\delta c_z^C \delta c_w^D \delta {\cal J}_y^B}; \qquad\quad \frac{\delta^2}{\delta c_z^{*C} \delta c_w^D}\left\langle \delta c_y^B\right\rangle = -i \varepsilon\cdot \frac{\delta^3\Gamma}{\delta c_z^{*C} \delta c_w^D \delta {\cal J}_y^B}.
\end{equation}

\noindent
In order to find explicit expressions for these Green functions, we note that the superfields $c$ and ${\cal J}$ are chiral. As a consequence, using the dimensional considerations we obtain that the corresponding contributions to the effective action in the momentum space can be written in the form

\begin{eqnarray}\label{H_Definition}
&& -\frac{e_0}{2} f^{ABC} \int d^2\theta \int \frac{d^4p}{(2\pi)^4} \frac{d^4q}{(2\pi)^4}\, c^A(\theta,q+p)\, c^B(\theta,-q)\, {\cal J}^C(\theta,-p)\, H(p,q)
\nonumber\\
&& + \frac{e_0}{8} f^{ABC} \int d^4\theta \int \frac{d^4p}{(2\pi)^4} \frac{d^4q}{(2\pi)^4}\, c^{*A}(\theta,q+p)\, c^B(\theta,-q)\, D^2 {\cal J}^C(\theta,-p)\, \widetilde H(p,q),\qquad
\end{eqnarray}

\noindent
where $H(p,q)$ is a dimensionless function and the function $\widetilde H(p,q)$ has the dimension $m^{-2}$. By construction, the function $H$ satisfies the relation

\begin{equation}
H(p,q) = H(p,-q-p).
\end{equation}

\noindent
Certainly, the functions $H$ and $\widetilde H$ also depend on the regularization parameter $\Lambda$ and the bare coupling constants, but, for simplicity, we do not write these arguments. From Eq. (\ref{H_Definition}) we conclude that the Green functions (\ref{Ghost_Green_Functions}) are explicitly written as

\begin{eqnarray}\label{Ghost_Function1}
&&\hspace*{-7mm} \frac{\delta^2}{\delta c_z^C \delta c_w^D}\left\langle \delta c_y^B\right\rangle = -\frac{i e_0 \varepsilon}{4} f^{BCD} \int \frac{d^4p}{(2\pi)^4} \frac{d^4q}{(2\pi)^4} H(p,q) \bar D_{z}^2\delta^8_{zy}(q+p) \bar D_{w}^2 \delta^8_{yw}(q);\\
&&\vphantom{1} \nonumber\\
\label{Ghostt_Function2}
&&\hspace*{-7mm} \frac{\delta^2}{\delta c_z^{*C} \delta c_w^D}\left\langle \delta c_y^B\right\rangle = -\frac{i e_0 \varepsilon}{64} f^{BCD} \int \frac{d^4p}{(2\pi)^4} \frac{d^4q}{(2\pi)^4} \widetilde H(p,q) \bar D_y^2 D_y^2\Big(D_z^2\delta^8_{zy}(q+p) \bar D_w^2 \delta^8_{yw}(q) \Big).
\end{eqnarray}

\noindent
Substituting these expressions into the Slavnov--Taylor identities (\ref{STI_Ghosts1}) and (\ref{STI_Ghosts2}) (and taking into account that the function $G_c$ actually depends on the square of the momentum) we can write them in the final form\footnote{For simplicity, we use the compact notation $G_c(-q^2/\Lambda^2) \to G_c(q)$. Also it should be noted that the scalar products of vectors are constructed using the Minkowski metric with the signature $(+---)$.}

\begin{eqnarray}\label{STI_For_Functions1}
&& G_c(q) F(q,p) + G_c(p) F(p,q) = 2 G_c(q+p) H(-q-p,q);\vphantom{\frac{1}{2}}\\
\label{STI_For_Functions2}
&& G_c(q) \widetilde F(q,p) - G_c(p) \Big(F(p,q) - 4 p^\mu F_\mu(p,q)\Big) = 2 G_c(q+p) (q+p)^2 \widetilde H(-q-p,q),\qquad\vphantom{\frac{1}{2}}
\end{eqnarray}

\noindent
where we take into account the identity $D^2 \bar D^2 D^2 = -16\partial^2 D^2$, which follows from the algebra of the supersymmetric covariant derivatives.

These identities can be easily verified in the tree approximation. Starting from the expression for the classical action (which includes the ghosts and the gauge fixing term) one can see that in this case

\begin{eqnarray}\label{Tree_Functions}
&& f(p,q) = O(\alpha_0,\lambda_0^2);\qquad F_{\mu}(p,q) = O(\alpha_0,\lambda_0^2);\qquad F(p,q) = \widetilde F(p,q) = 1 + O(\alpha_0,\lambda_0^2);\qquad
\vphantom{\frac{1}{2}} \nonumber\\
&& \qquad G_c(q) = 1 + O(\alpha_0,\lambda_0^2); \qquad H(p,q) = 1+ O(\alpha_0,\lambda_0^2);\qquad \widetilde H(p,q) = O(\alpha_0,\lambda_0^2).\qquad\vphantom{\frac{1}{2}}
\end{eqnarray}

\noindent
Really, for example, a part of the classical action corresponding to the $\bar c^* V c$-vertex has the form

\begin{equation}
\frac{1}{2 e_0^2} \mbox{tr} \int d^8x\, \bar c^+ [V, c] = \frac{ie_0}{4} f^{ABC} \int d^8x\, c^{*A} V^B c^C.
\end{equation}

\noindent
Differentiating this expression we obtain

\begin{equation}
\frac{\delta^3S_{\mbox{\scriptsize FP}}}{\delta \bar c_x^{*A} \delta V_y^B \delta c_z^C} = -\frac{ie_0}{16} f^{ABC} D_x^2 \delta^8_{xy}\, \bar D_z^2 \delta^8_{yz}.
\end{equation}

\noindent
This implies that in the tree approximation $F=1$, $F_\mu=0$, and $f=0$. The other functions are constructed in a similar way.

Substituting the expressions (\ref{Tree_Functions}) into Eqs. (\ref{STI_For_Functions1}) and (\ref{STI_For_Functions2}) it can be easily verified that the Slavnov--Taylor identities are really valid in this approximation. The one-loop verification of these identities will be presented below.

\section{Non-renormalization of the $V\bar c c$ vertices}
\hspace*{\parindent}\label{Section_Non_Renormalization}

In order to prove the non-renormalization theorem for the $V\bar c c$-vertices we first consider the structure of quantum corrections to the function $H(p,q)$ which is defined by Eq. (\ref{Ghost_Function1}). These quantum corrections are given by diagrams in which one leg corresponds to the chiral source ${\cal J}$ and two other legs correspond to the chiral ghost superfields $c$. Let us consider an arbitrary supergraph of this type (with an arbitrary number of loops) and denote the vertex containing the source ${\cal J}$ by $y$. This vertex also contains the product of two ghost propagators with $y$ chiral ends (their antichiral ends we denote by the subscripts $1$ and $2$):

\begin{equation}
\int d^4y\, d^2\theta_y\, {\cal J}_y^A \cdot \frac{\bar D_y^2 D_y^2}{4\partial^2} \delta^8_{y1}\cdot \frac{\bar D_y^2 D_y^2}{4\partial^2} \delta^8_{y2} = -2\int d^4y\, d^4\theta_y\, {\cal J}_y^A \cdot \frac{D_y^2}{4\partial^2} \delta^8_{y1}\cdot \frac{\bar D_y^2 D_y^2}{4\partial^2} \delta^8_{y2}.
\end{equation}

\noindent
(This structure of the vertex follows from Eq. (\ref{Ghost_Source}).) Using the standard technique for calculating supergraphs \cite{West:1990tg,Buchbinder:1998qv} we see that the considered contribution to the effective action is given by an integral over the whole superspace. In particular, this implies that it includes integration over $d^4\theta$. From the other side, any loop correction to the considered correlator is presented in the form of the first term of Eq. (\ref{H_Definition}). Taking into account that

\begin{equation}
\int d^4\theta = -\frac{1}{2} \int d^2\theta \bar D^2 + \mbox{total derivatives in the coordinate space},
\end{equation}

\noindent
we obtain that two left spinor derivatives should act to the external lines. However, the external lines are chiral. Therefore, according to the standard rules for supergraph calculating, a non-trivial result can be obtained only if two right spinor derivatives also act to the external lines. Thus, we conclude that the result should be proportional to, at least, the second degree of the external momenta. Therefore, the corresponding integrals are proportional to $\Lambda^{-2}$ and do not contain ultraviolet divergencies. This implies that the function $H(p,q)$ is UV finite. (Due to the renormalizability, the UV subdivergencies are also absent if the function $H$ is written in terms of the renormalized coupling constants.) Below we will demonstrate the finiteness of the function $H(p,q)$ by the explicit calculation in the one-loop approximation.

Let us construct the renormalization constant $Z_c$ which is defined so that the renormalized Green function

\begin{equation}
(G_c)_{R}(\alpha,\lambda,q^2/\mu^2) = \lim\limits_{\Lambda\to \infty} Z_c(\alpha,\lambda,\Lambda/\mu)\, G_c(\alpha_0,\lambda_0,q^2/\Lambda^2)
\end{equation}

\noindent
is finite in the UV region. Here $\mu$ is the renormalization point, $\alpha$ and $\lambda$ are the renormalized coupling and Yukawa constants, respectively. By construction, $(G_c)_{R}$ does not depend on the parameter $\Lambda$ in the higher covariant derivative term.

In order to prove the non-renormalization of the $V\bar c c$-vertex we multiply the Slavnov--Taylor identity (\ref{STI_For_Functions1}) by the renormalization constant $Z_c$, and express both sides of this equation in terms of the renormalized coupling constants. After this, we make the differentiation with respect to $\ln\Lambda$ at fixed values of the renormalized coupling constants and take the limit $\Lambda\to \infty$. Due to the finiteness of the renormalized ghost two-point Green function $(G_c)_{\mbox{\scriptsize ren}}$ and of the function $H$ (expressed in terms of the renormalized coupling constants) the right hand side vanishes and we obtain

\begin{equation}
\Big((G_c)_{R}(q) \frac{d}{d\ln\Lambda} F(q,p) + (G_c)_{R}(p) \frac{d}{d\ln\Lambda} F(p,q)\Big)\Bigg|_{\Lambda\to \infty} = 0.
\end{equation}

\noindent
Next, we set in this equation $p=-q$ and take into account that the function $G_c$ depends only on the squared momentum. Therefore,

\begin{equation}
\frac{d}{d\ln\Lambda} \Big(F(-q,q) + F(q,-q)\Big)\Bigg|_{\Lambda\to \infty} = 0,
\end{equation}

\noindent
where the derivative with respect to $\ln\Lambda$ should be calculated at fixed values of the renormalized coupling and Yukawa constants. Because the function $F(-q,q)$ depends only on $\ln(\Lambda^2/q^2)$, we conclude that it is finite. Therefore, the corresponding renormalization constant (see below) is finite. As a consequence, the function $F(p,q)$ is also finite.

Terms in the effective action corresponding to the Green function (\ref{Three_Point_Function1}) have the form

\begin{eqnarray}\label{Three_Point_Contribution}
&& \frac{i e_0}{4} f^{ABC} \int d^4\theta\, \frac{d^4p}{(2\pi)^4} \frac{d^4q}{(2\pi)^4} \bar c^{*A}(\theta,p+q)\Big(f(p,q) \partial^2\Pi_{1/2}V^B(\theta,-p)\nonumber\\
&&\qquad\qquad\qquad\quad + F_\mu(p,q) (\gamma^\mu)_{\dot a}{}^{b} D_b \bar D^{\dot a} V^B(\theta,-p) + F(p,q) V^B(\theta,-p)\Big) c^C(\theta,-q).\qquad
\end{eqnarray}

\noindent
Writing this expression in terms of the renormalized values according to Eq. (\ref{Z_Definition}) we conclude that the renormalized function $F$ is given by

\begin{equation}\label{F_R}
F_R(p,q) = Z_\alpha^{-1/2} Z_c Z_V F(p,q).
\end{equation}

\noindent
Note that deriving this equation we take into account the presence of the bare coupling constant $e_0$ in Eq. (\ref{Three_Point_Contribution}). Similar equations can be also written for the functions $f(p,q)$ and $F_\mu(p,q)$. From Eq. (\ref{F_R}) we see that the finiteness of the function $F$ leads to the relation

\begin{equation}
\frac{d}{d\ln\Lambda} (Z_\alpha^{-1/2} Z_c Z_V) = 0.
\end{equation}

\noindent
Although the renormalization constants are not uniquely defined, it is possible to choose the subtraction scheme in which

\begin{equation}\label{Z_Relation}
-\frac{1}{2} \ln Z_\alpha + \ln Z_c + \ln Z_V  = 0.
\end{equation}

Finally, we note that the renormalization constants for all 4 vertices of the type $V \bar c c$, i.e., proportional to $\bar c [V, c]$, $\bar c^+[V, c]$, $\bar c [V, c^+]$, and $\bar c^+ [V,c^+]$ are the same. Therefore, finiteness of the vertex $\bar c^+ [V, c]$, which was proved above, leads to finiteness of the other vertices. Certainly, this is a consequence of the renormalizability of ${\cal N}=1$ gauge supersymmetric theories.

\section{One-loop verification}
\hspace*{\parindent}\label{Section_One_Loop}

It is desirable to verify the proof made in the previous section using a rather complicated technique by explicit calculations in the lowest loops. In this section we make such a verification in the one-loop approximation. Namely, in the considered approximation we calculate the Faddeev-Popov ghost two-point function $G_c$, the three-point $\bar c\, V c$-vertices, and the Green functions (\ref{Ghost_Green_Functions}). Next, we check the Slavnov--Taylor identities (verifying, thereby, correctness of the calculation) and demonstrate finiteness of the function $H$ and $\bar c\, V c$-vertices.

\begin{figure}[h]
\begin{picture}(0,1.9)
\put(4.1,0.1){\includegraphics[scale=0.42]{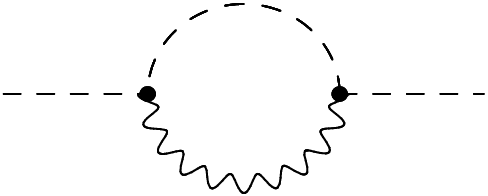}}
\put(4.1,1.0){$\bar c^*$} \put(7.4,1.0){$c$}
\put(8.7,-0.1){\includegraphics[scale=0.41]{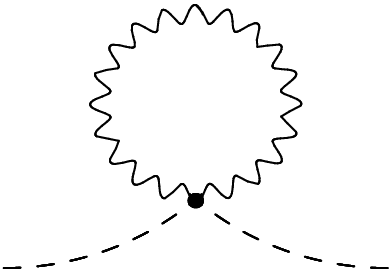}}
\put(8.6,0.1){$\bar c^*$} \put(11.3,0.1){$c$}
\end{picture}
\caption{Diagrams contributing to the Faddeev--Popov ghost two-point function.}\label{Figure_Ghost_Function}
\end{figure}
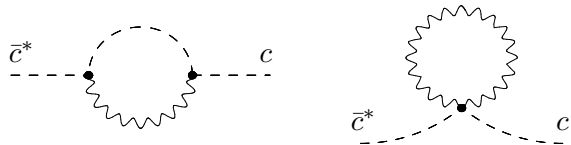

The function $G_c(p)$ is defined by Eq. (\ref{G_C_Definition}) so that in the tree approximation $G_c=1$. The one-loop correction to this function is determined by two diagrams presented in Fig. \ref{Figure_Ghost_Function}. Having calculated these diagrams we obtained

\begin{equation}\label{One_Loop_G_c}
\qquad G_c(p) = 1 + e_0^2 C_2 \int \frac{d^4k}{(2\pi)^4} \Big(\frac{\xi_0}{K_k} - \frac{1}{R_k}\Big) \Big(- \frac{1}{6k^4} + \frac{1}{2k^2 (k+p)^2} - \frac{p^2}{2k^4 (k+p)^2}\Big)  + O(e_0^4, e_0^2 \lambda_0^2),\qquad
\end{equation}

\noindent
where $R_k \equiv R(k^2/\Lambda)$ and $K_k \equiv K(k^2/\Lambda^2)$. This expression is written in the Euclidean space after the Wick rotation as function of the Euclidean momentum $p^\mu$ with $p_4 = -ip_0$. From the expression (\ref{One_Loop_G_c}) we see that the considered function is divergent in the ultraviolet region (for infinite $\Lambda$). Certainly, for finite $\Lambda$ the integral is UV finite due to the higher derivatives in the denominator (inside the functions $R$ and $K$). It is should be noted that the integral is divergent in the IR region. Such divergences are well-known \cite{Itzykson:1980rh}. Usually, they are regularized by the substitution $k^2 \to k^2+m^2$, where $m$ is a small dimensionful parameter. However, in this paper we are interested only in the ultraviolet divergences and will ignore the infrared effects. It is sufficient to point out that well-defined expressions can be obtained by differentiating with respect to $\ln\Lambda$ and taking the limit $\Lambda\to \infty$:

\begin{equation}
\gamma_c(\alpha_0,\lambda_0) = \left. \frac{d\ln G_c}{d\ln\Lambda}\right|_{p=0;\,\alpha,\lambda=\mbox{\scriptsize const}} = - \frac{\alpha_0 C_2
(1-\xi_0)}{6\pi} + O(\alpha_0^2,\alpha_0\lambda_0^2).
\end{equation}

\noindent
(This result is in agreement with the calculation made in \cite{Aleshin:2016yvj}.)

\begin{figure}[h]
\begin{picture}(0,5.6)
\put(2.1,3.1){\includegraphics[scale=0.16]{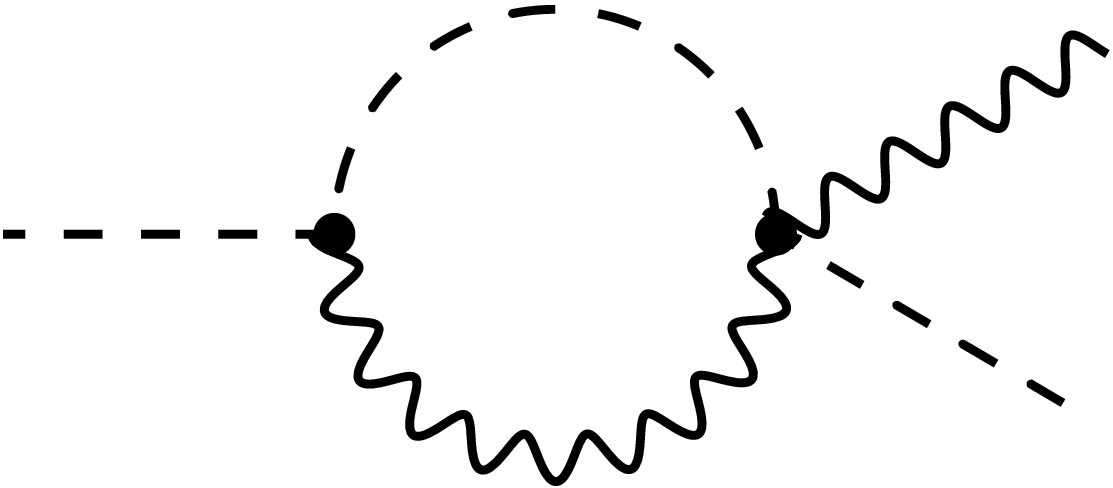}}
\put(6.6,3.1){\includegraphics[scale=0.16]{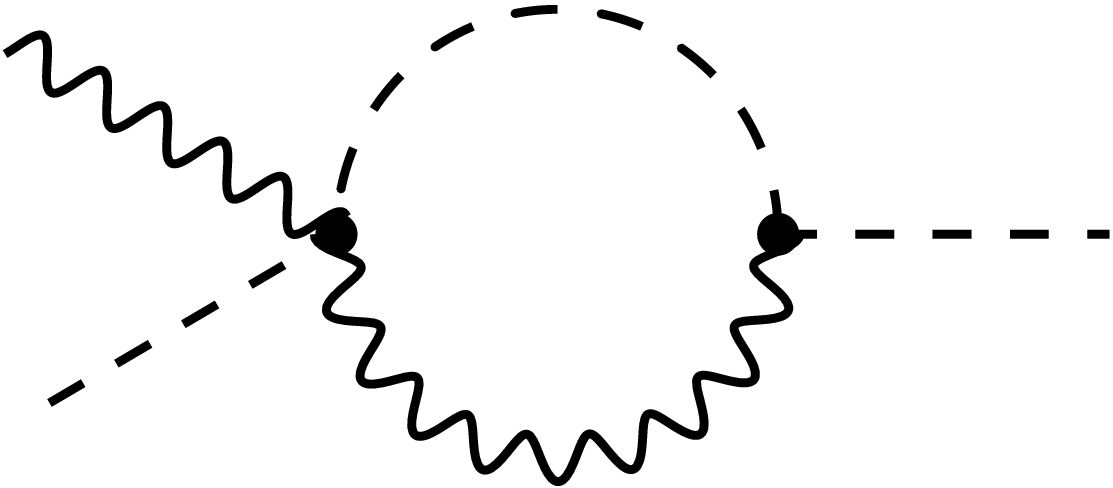}}
\put(11.0,3.08){\includegraphics[scale=0.16]{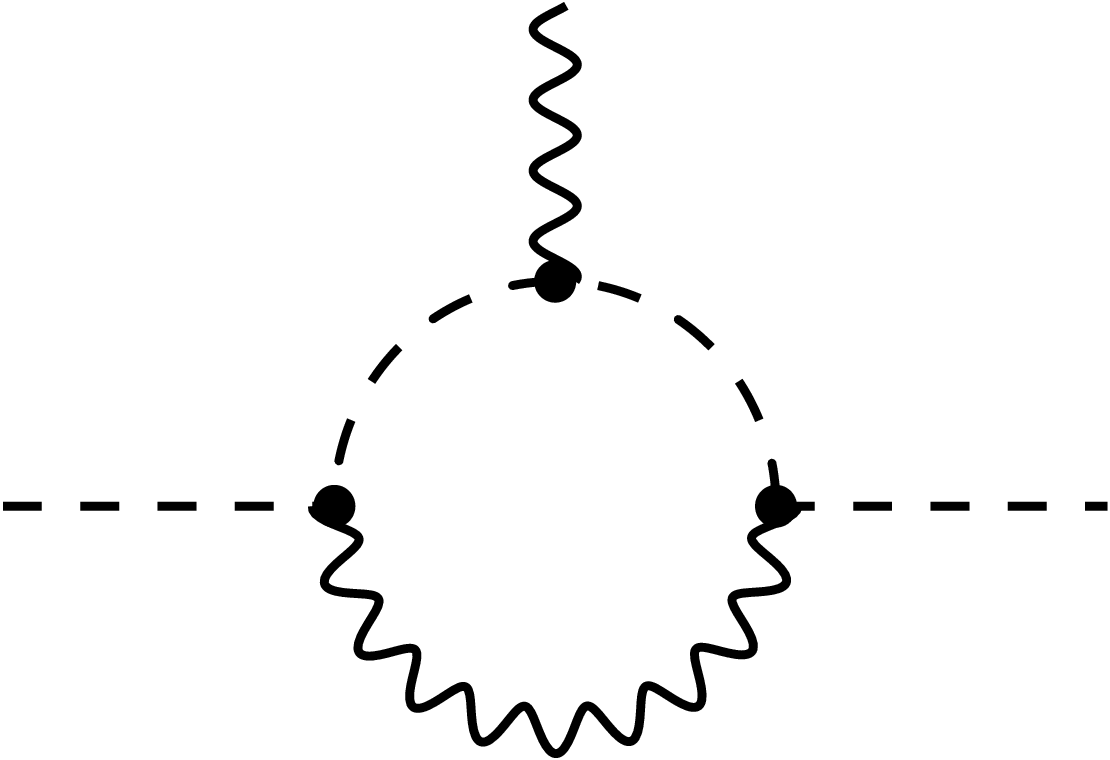}}
\put(2.1,0.0){\includegraphics[scale=0.16]{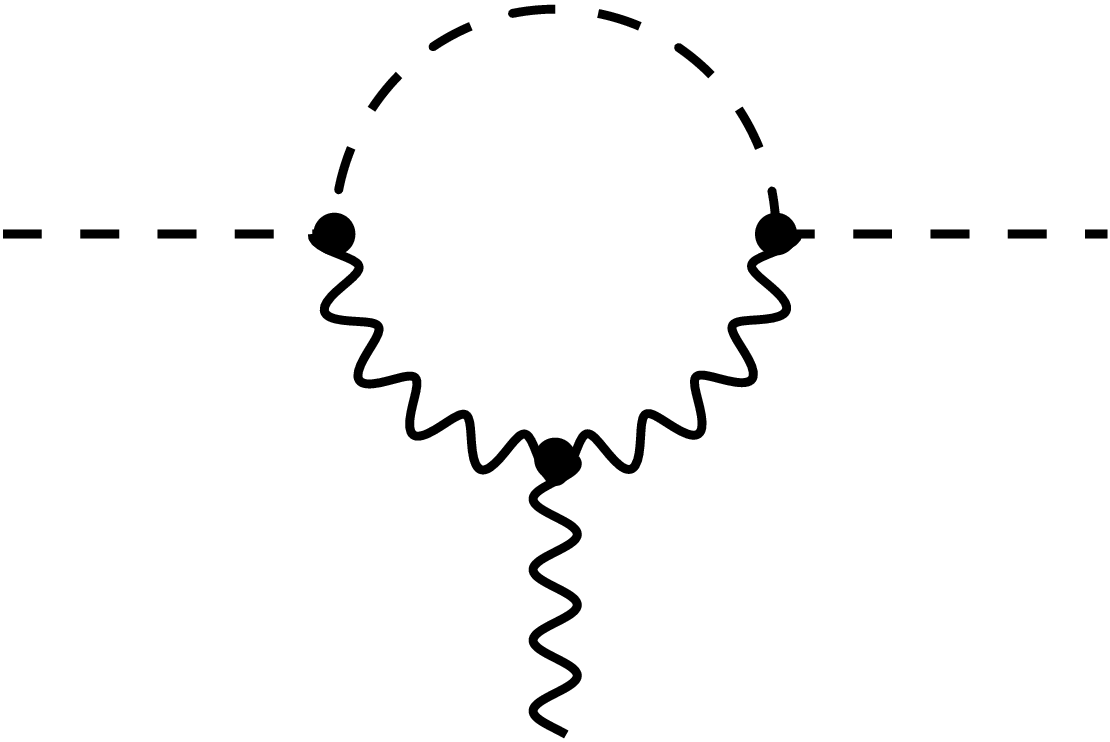}}
\put(6.9,0.2){\includegraphics[scale=0.153]{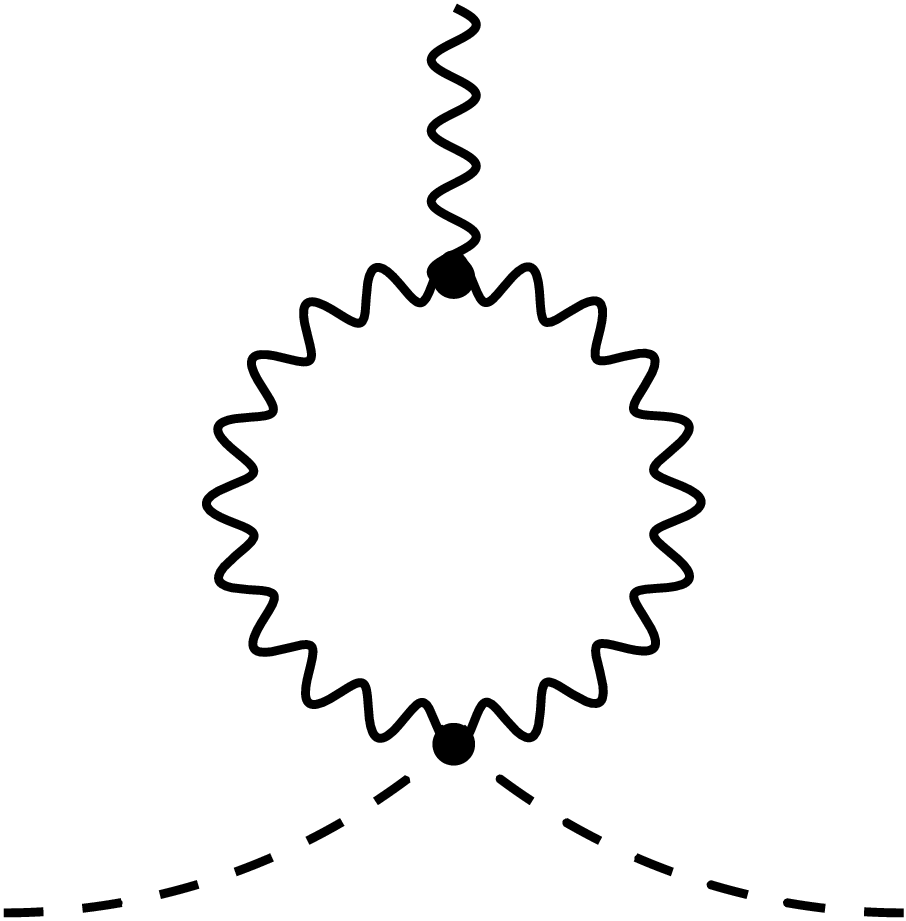}}
\put(11.3,-0.15){\includegraphics[scale=0.153]{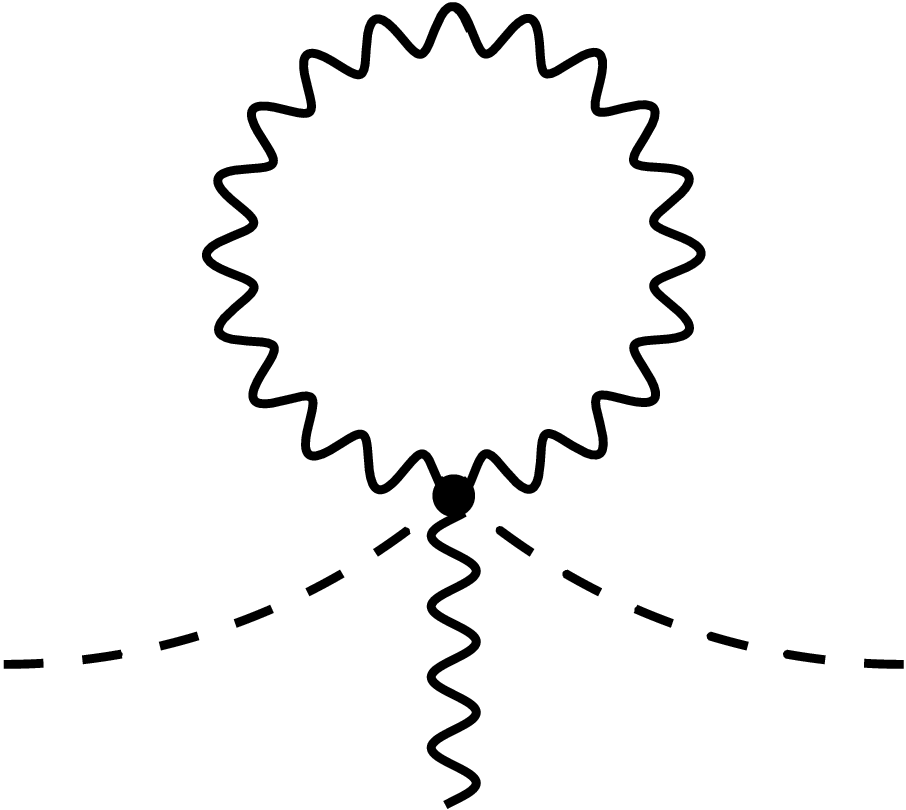}}
\put(2.0,3.95){$\bar c^*$} \put(5.1,4.5){$V$}  \put(4.8,2.9){$c$ or $c^*$}
\end{picture}
\caption{One-loop graphs contributing to the three-point Green functions $\bar c^* V c$ and $\bar c^* V c^*$. In all these diagrams the left end of the dashed line corresponds to the superfield $\bar c^*$ and the right end of the dashed line corresponds to the superfields $c$ or $c^*$. The wavy external line corresponds to the quantum gauge superfield $V$.}\label{Figure_VCC_Function}
\end{figure}
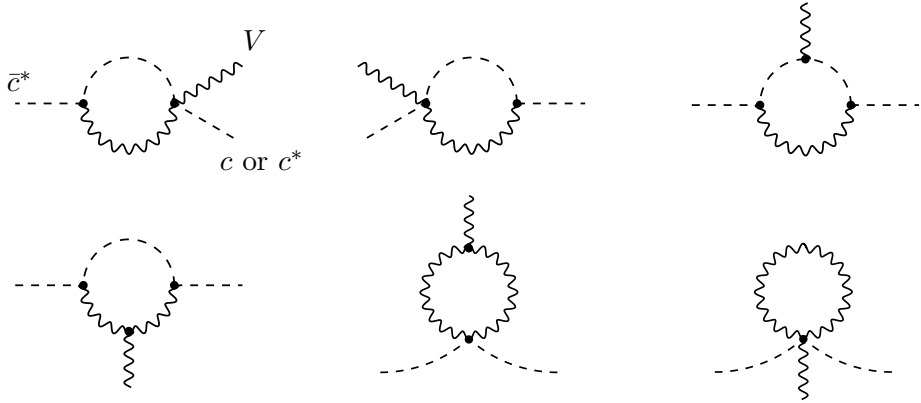

The three-point $V\bar c c$ Green functions in the one-loop approximation are determined by the diagrams presented in Fig. \ref{Figure_VCC_Function}. After calculating these diagrams we have obtained the one-loop results for the functions entering Eqs. (\ref{Three_Point_Function1}) and (\ref{Three_Point_Function2}). Here we present the results for the functions $F(p,q)$, $\widetilde F(p,q)$ of the Euclidean momentums $p$ and $q$:

\begin{eqnarray}
&& F(p,q) = 1 + \frac{e_0^2 C_2}{4} \int \frac{d^4k}{(2\pi)^4} \Bigg\{-\frac{(q+p)^2}{R_k k^2 (k+p)^2 (k-q)^2} - \frac{\xi_0\, p^2}{K_k k^2 (k+q)^2 (k+q+p)^2}\nonumber\\
&& + \frac{\xi_0\, q^2}{K_k k^2 (k+p)^2 (k+q+p)^2} + \left(\frac{\xi_0}{K_k} - \frac{1}{R_k}\right)\left(- \frac{2(q+p)^2}{k^4 (k+q+p)^2}  + \frac{2}{k^2 (k+q+p)^2} \right.\nonumber\\
&&\left. -\frac{1}{k^2 (k+q)^2} - \frac{1}{k^2 (k+p)^2} \right)\Bigg\}+ O(\alpha_0^2,\alpha_0\lambda_0^2).
\end{eqnarray}

\begin{eqnarray}
&& \widetilde F(p,q) = 1 - \frac{e_0^2 C_2}{4} \int \frac{d^4k}{(2\pi)^4} \Bigg\{\frac{p^2}{R_k k^2 (k+q)^2 (k+q+p)^2}
+ \frac{\xi_0\, (q+p)^2}{K_k k^2 (k-p)^2 (k+q)^2}\quad\ \nonumber\\
&& + \frac{\xi_0\, q^2}{K_k k^2 (k+p)^2 (k+q+p)^2} +\frac{2\xi_0}{K_k k^2 (k+p)^2} -\frac{2\xi_0}{K_k k^2(k+q+p)^2} + \left(\frac{\xi_0}{K_k} - \frac{1}{R_k}\right)
\nonumber\\
&& \times \left( \frac{2q^2}{k^4 (k+q)^2}  + \frac{1}{k^2 (k+q+p)^2} -\frac{1}{k^2 (k+q)^2} \right)\Bigg\}+ O(\alpha_0^2,\alpha_0\lambda_0^2).
\end{eqnarray}

\noindent
From these expressions we see that the considered functions are really finite in the limit $\Lambda\to\infty$. The remaining functions $f(p,q)$ and $F_\mu(p,q)$
are also finite in the one-loop approximation, because they have the dimensions $m^{-2}$ and $m^{-1}$, respectively. The expressions for them are much larger, and we explicitly write them in the Appendix. Certainly, their finiteness can be easily seen from the explicit expressions presented there.

\begin{figure}[h]
\begin{picture}(0,2.6)
\put(3.1,0.0){\includegraphics[scale=0.19]{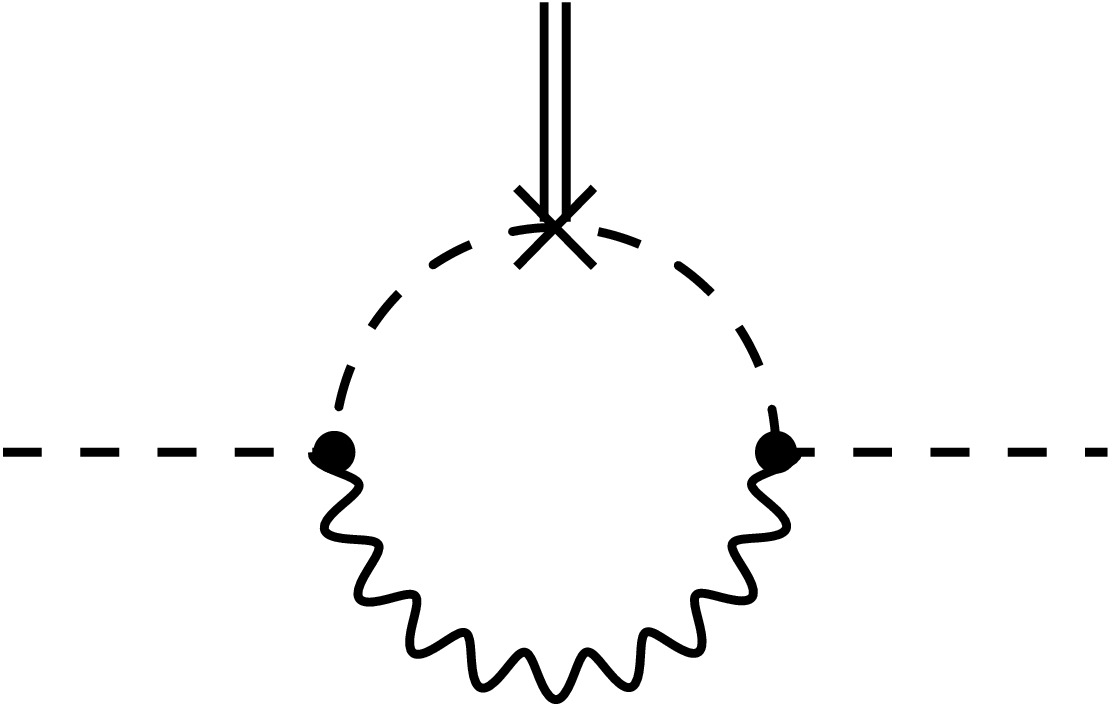}}
\put(3.1,1.0){$c$} \put(6.5,1.0){$c$}  \put(5.1,2.3){${\cal J}$} \put(4.45,1.6){$c$} \put(5.15,1.6){$c$} \put(3.83,1.0){$\bar c^*$} \put(5.7,1.0){$\bar c^*$}
\put(9.1,0.0){\includegraphics[scale=0.19]{jcc.eps}}
\put(9.05,1.0){$c^*$} \put(12.5,1.0){$c$}  \put(11.1,2.3){${\cal J}$} \put(10.45,1.6){$c$} \put(11.15,1.6){$c$} \put(9.83,1.0){$\bar c^*$} \put(11.7,1.0){$\bar c^*$}
\end{picture}
\caption{The graph from the left and the graph from the right determine the one-loop quantum corrections to the functions $H(p,q)$ and $\widetilde H(p,q)$, respectively.}\label{Figure_H_Function}
\end{figure}

In order to verify correctness of the calculations and the general arguments presented in the previous section it is necessary to check the Slavnov--Taylor identities (\ref{STI_For_Functions1}) and (\ref{STI_For_Functions2}). Taking into account the relation between Minkowski and Euclidean scalar products, $(a_\mu b^\mu)_M = - (a_\mu b^\mu)_E$, and using the result for the function $F_\mu(p,q)$ it is possible to obtain\footnote{Note that this expression is written as a function of the Euclidean momentums.}

\begin{eqnarray}
&& F(p,q) + 4 p^\mu F_\mu(p,q) = 1 + \frac{e_0^2 C_2}{4} \int \frac{d^4k}{(2\pi)^4} \Bigg\{\frac{2}{R_k k^2 (k+q+p)^2} - \frac{2}{R_k k^2 (k+q)^2}\nonumber\\
&& -\frac{\xi_0 p^2}{K_k k^2 (k+q)^2 (k+q+p)^2} -\frac{q^2}{R_k k^2 (k+p)^2 (k+q+p)^2} + \frac{\xi_0 (q+p)^2}{K_k k^2 (k-p)^2 (k+q)^2}\nonumber\\
&& + \left(\frac{\xi_0}{K_k} - \frac{1}{R_k}\right)\left(-\frac{2q^2}{k^4(k+q)^2}
- \frac{1}{k^2 (k+p)^2} + \frac{1}{k^2 (k+q+p)^2}\right)\Bigg\} + O(\alpha_0^2,\alpha_0\lambda_0^2).\qquad
\end{eqnarray}

\noindent
Moreover, it is necessary to calculate one-loop contributions to the functions $H(p,q)$ and $\widetilde H(p,q)$ which are defined by Eq. (\ref{H_Definition}). For this purpose we calculate the diagrams presented in Fig. \ref{Figure_H_Function}. In these diagrams two external lines correspond to the chiral ghosts superfields $c$ and the double external line corresponds to the chiral source ${\cal J}$. After calculating this graph we obtained that the sums of the tree and one-loop contributions have the form

\begin{eqnarray}
&& H(p,q) = 1 - \frac{e_0^2 C_2}{4} \int \frac{d^4k}{(2\pi)^4} \Bigg\{\frac{p^2}{R_k k^2 (k+q)^2 (k+q+p)^2}\nonumber\\
&&\qquad\qquad\quad + \frac{(q+p)^2}{k^4 (k+q+p)^2} \Big(\frac{\xi_0}{K_k} - \frac{1}{R_k}\Big) + \frac{q^2}{k^4 (k+q)^2} \Big(\frac{\xi_0}{K_k} - \frac{1}{R_k}\Big) \Bigg\} + O(e_0^4, e_0^2 \lambda_0^2);\qquad\\
&&\vphantom{1}\nonumber\\
&& \widetilde H(p,q) = \frac{e_0^2 C_2}{4} \int \frac{d^4k}{(2\pi)^4} \frac{\xi_0}{K_k k^2 (k+q)^2 (k+q+p)^2} + O(e_0^4, e_0^2 \lambda_0^2).
\end{eqnarray}

\noindent
We see that the function $H$ is finite in the ultraviolet region and proportional to the second degree of external momenta, as it was argued in the previous section. (Let us remind that the ultraviolet finiteness of the functions $H(p,q)$ is a key ingredient of $V\bar c c$-non-renormalization theorem proof.) The function $\widetilde H$ is UV finite and in the considered approximation does not also contain IR divergencies. However, the infrared divergences are present in the expression for the function $H$. The well-defined expression is obtained after differentiating with respect to $\ln\Lambda$ (at fixed values of the renormalized coupling constants) and subsequent taking the limit $\Lambda\to \infty$:

\begin{equation}
\left.\frac{dH(p,q)}{d\ln\Lambda}\right|_{\alpha,\lambda=\mbox{\scriptsize const};\,\Lambda\to\infty} = 0.
\end{equation}

Using the above expressions one can easily verify the Slavnov--Taylor identities. For example, in the one-loop approximation Eq. (\ref{STI_For_Functions1}) gives

\begin{eqnarray}
&&\hspace*{-6mm} G_c(-q-p) H(-q-p,q) = 1 + \frac{e_0^2 C_2}{4} \int \frac{d^4k}{(2\pi)^4} \Bigg\{-\frac{(q+p)^2}{R_k k^2 (k+p)^2 (k-q)^2} + \Big(\frac{\xi_0}{K_k} - \frac{1}{R_k}\Big)\nonumber\\
&&\hspace*{-6mm} \times \left(\frac{2}{k^2 (k+q+p)^2} - \frac{2(q+p)^2}{k^4 (k+q+p)^2}-\frac{p^2}{k^4 (k+p)^2} - \frac{q^2}{k^4 (k-q)^2} -\frac{2}{3 k^4}\right)\Bigg\} + O(\alpha_0^2,\alpha_0\lambda_0^2)\nonumber\\
&&\hspace*{-6mm} = \frac{1}{2}\Big(G_c(q) F(q,p) + G_c(p) F(p,q)\Big).\vphantom{\Bigg\{}
\end{eqnarray}

\noindent
The identity (\ref{STI_For_Functions2}) is verified by the similar way:\footnote{Let us remind that in Eq. (\ref{STI_For_Functions2}) we use Minkowski momentums, while here the momentums are Euclidean. Thus, due to the identity $(a_\mu b^\mu)_M = - (a_\mu b^\mu)_E$ some signs are different.}

\begin{eqnarray}
&&\hspace*{-6mm} G_c(q) \widetilde F(q,p) - G_c(p) \Big(F(p,q) + 4 p^\mu F_\mu(p,q)\Big) = - \frac{e_0^2 C_2}{2}  \int \frac{d^4k}{(2\pi)^4} \frac{\xi_0(q+p)^2}{K_k k^2 (k-p)^2 (k+q)^2}\quad\nonumber\\
&&\hspace*{-6mm} + O(\alpha_0^2, \alpha_0 \lambda_0^2) = - 2 G_c(q+p) (q+p)^2 \widetilde H(-q-p,q).\vphantom{\frac{1}{2}}
\end{eqnarray}

\noindent
Therefore, all steps of the non-renormalization theorem proof are confirmed by the explicit one-loop calculation.

\section{NSVZ $\beta$-function in terms of $\gamma_c$ and $\gamma_V$}
\hspace*{\parindent}\label{Section_NSVZ}

The NSVZ $\beta$-function written in terms of the bare charges has the following form:

\begin{equation}\label{NSVZ}
\beta(\alpha_0,\lambda_0) = - \frac{\alpha_0^2\Big(3 C_2 - T(R) + C(R)_i{}^j (\gamma_\phi)_j{}^i(\alpha_0,\lambda_0)/r\Big)}{2\pi(1-C_2\alpha_0/2\pi)},
\end{equation}

\noindent
where $\alpha_0$ and $\lambda_0$ are bare coupling and Yukawa constants, respectively, and the constants entering this equation are defined by

\begin{eqnarray}
&& \mbox{tr}\,(T^A T^B) \equiv T(R)\,\delta^{AB};\qquad
(T^A)_i{}^k
(T^A)_k{}^j \equiv C(R)_i{}^j;\qquad\ \nonumber\\
&& f^{ACD} f^{BCD} \equiv C_2 \delta^{AB};\qquad\quad r \equiv
\delta_{AA}.\qquad
\end{eqnarray}

\noindent
Eq. (\ref{NSVZ}) can be equivalently rewritten as

\begin{equation}\label{NSVZ_Auxiliary}
\frac{\beta(\alpha_0,\lambda_0)}{\alpha_0^2} = - \frac{3 C_2 - T(R) + C(R)_i{}^j (\gamma_\phi)_j{}^i(\alpha_0,\lambda_0)/r}{2\pi}
+ \frac{C_2}{2\pi}\cdot \frac{\beta(\alpha_0,\lambda_0)}{\alpha_0}.
\end{equation}

\noindent
The $\beta$-function can be expressed in terms of the renormalization constant $Z_\alpha$ as

\begin{equation}
\beta(\alpha_0,\lambda_0) = \frac{d\alpha_0(\alpha,\lambda,\Lambda/\mu)}{d\ln\Lambda}\Big|_{\alpha,\lambda=\mbox{\scriptsize const}}
= -\alpha_0 \frac{d\ln Z_\alpha}{d\ln\Lambda}\Big|_{\alpha,\lambda=\mbox{\scriptsize const}}.
\end{equation}

\noindent
Eq. (\ref{Z_Relation}) allows relating $Z_\alpha$ to the renormalization constants $Z_c$ and $Z_V$. Therefore, due to the non-renormalization theorem for the $V\bar cc$ vertex it is possible to present the $\beta$-function (defined in terms of the bare charges) in the form

\begin{equation}
\beta(\alpha_0,\lambda_0)
= -2\alpha_0 \frac{d\ln (Z_c Z_V)}{d\ln\Lambda}\Big|_{\alpha,\lambda=\mbox{\scriptsize const}} = 2\alpha_0 \Big(\gamma_c(\alpha_0,\lambda_0) + \gamma_V(\alpha_0,\lambda_0)\Big),
\end{equation}

\noindent
where $\gamma_c$ and $\gamma_V$ are anomalous dimensions of the Faddeev--Popov ghosts and of the quantum gauge superfield, respectively.
Substituting this expression to the right hand side of Eq. (\ref{NSVZ_Auxiliary}) one can rewrite the NSVZ $\beta$-function as

\begin{equation}\label{NSVZ_With_Gamma}
\frac{\beta(\alpha_0,\lambda_0)}{\alpha_0^2} = - \frac{1}{2\pi}\Big(3 C_2 - T(R) - 2C_2 \gamma_c(\alpha_0,\lambda_0) - 2C_2 \gamma_V(\alpha_0,\lambda_0) + C(R)_i{}^j (\gamma_\phi)_j{}^i(\alpha_0,\lambda_0)/r\Big).
\end{equation}

\begin{figure}[h]
\begin{picture}(0,3.5)
\put(5,1){\includegraphics[scale=0.18]{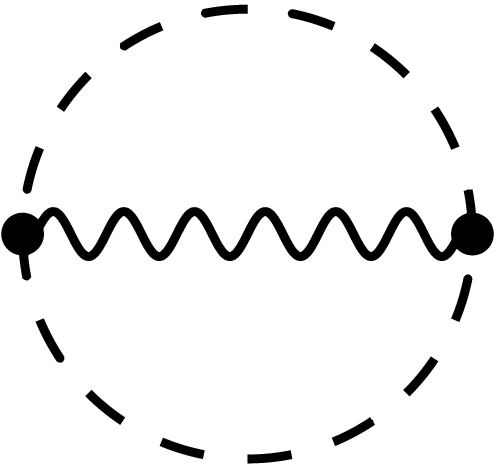}}
\put(7,1){\includegraphics[scale=0.5,angle=-15]{arrow.eps}}
\put(7,2){\includegraphics[scale=0.5,angle=15]{arrow.eps}}
\put(9,0){\includegraphics[scale=0.4]{ghost_z1}}
\put(9,2){\includegraphics[scale=0.4]{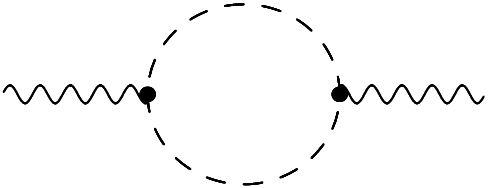}}
\end{picture}
\caption{Cutting the graph without external lines in the non-Abelian case gives diagrams contributing to the anomalous dimensions of the quantum gauge superfield and of the Faddeev--Popov ghosts.}\label{Figure_NSVZ_Derivation}
\end{figure}

\noindent
This form seems to be convenient for deriving the NSVZ relation by the direct summation of Feynman diagrams, because the Faddeev--Popov ghosts and the ordinary matter superfields enter in the right hand side in a similar way. Moreover, the calculations made in the Abelian case and for the Adler $D$-function in ${\cal N}=1$ SQCD reveal the qualitative picture of appearing the NSVZ expression in the perturbation theory in the case of using the higher covariant derivative regularization, which is illustrated by Fig. \ref{Figure_Qualitative_Picture}. Namely, we start with a graph without external lines. Then the contribution to the $\beta$-function is obtained after calculating all diagrams which are obtained from the original graph be attaching two external lines of the background gauge superfield. From the other side, the diagrams corresponding to the right hand side of the NSVZ relation are obtained by various cutting of the original graph. Certainly, as a result of this procedure in the non-Abelian case we will obtain diagrams contributing to the anomalous dimensions of the matter superfields, Faddeev--Popov ghosts, and quantum gauge superfield. The example is presented in Fig. \ref{Figure_NSVZ_Derivation}.

Exactly these anomalous dimensions are present in the right hand side of Eq. (\ref{NSVZ_With_Gamma}). That is why we believe that it is Eq. (\ref{NSVZ_With_Gamma}) that will be obtained by summing the diagrams in the perturbation theory. More exactly, taking into account the results of \cite{Aleshin:2016yvj}, the derivation of the NSVZ relation in the non-Abelian case for the RG functions defined in terms of the bare charges should be made by proving the following relation between the Green functions

\begin{eqnarray}\label{NSVZ_Consequence}
&& \frac{d}{d\ln\Lambda}\Big(d^{-1} - \alpha_0^{-1}\Big)\Big|_{\alpha,\lambda=\mbox{\scriptsize const};\ p\to 0}
= - \frac{3 C_2 - T(R)}{2\pi}\nonumber\\
&&\qquad\quad - \frac{1}{2\pi}\frac{d}{d\ln\Lambda}\Big(- 2C_2 \ln G_c - C_2 \ln G_V + C(R)_i{}^j \ln (G_\phi)_j{}^i(\alpha_0,\lambda_0)/r\Big)\Big|_{\alpha,\lambda=\mbox{\scriptsize const}; q\to 0}.\qquad
\end{eqnarray}

\noindent
The functions $d^{-1}$, $(G_\phi)_i^j$, $G_c$, and $G_V$ are constructed according to the prescription

\begin{eqnarray}
&& \Gamma^{(2)} - S^{(2)}_{\mbox{\scriptsize gf}} = \frac{1}{4} \int
\frac{d^4p}{(2\pi)^4}\, d^4\theta\, \phi^{*i}(\theta,-p)
\phi_j(\theta,p) (G_\phi)_i{}^j(\alpha_0,\lambda_0,\Lambda/p) +\mbox{tr} \int
\frac{d^4p}{(2\pi)^4}\,d^4\theta\,\nonumber\\
&& \times \Big[- \frac{1}{8\pi} \bm{V}(\theta,-p)\,\partial^2\Pi_{1/2}
\bm{V}(\theta,p)\,
d^{-1}(\alpha_0,\lambda_0,\Lambda/p) - \frac{1}{2e_0^2}\,V(\theta,-p)\,\partial^2\Pi_{1/2}
V(\theta,p)\,\\
&& \times G_V(\alpha_0,\lambda_0,\Lambda/p) + \frac{1}{2e_0^2}\,\Big(-\bar c(\theta,-p) c^+(\theta,p) + \bar
c^+(\theta,-p) c(\theta,p) \Big)
G_c(\alpha_0,\lambda_0,\Lambda/p)\Big]+\ldots,\qquad\nonumber
\end{eqnarray}

\noindent
where dots denote the other possible quadratic contributions.

Equation (\ref{NSVZ_Consequence}) can be easily derived from Eq. (\ref{NSVZ_With_Gamma}) using finiteness of the renormalized Green functions $Z_c G_c$, $Z_V^2 G_V$, and $Z_i{}^j G_j{}^k$. The external momentums $p$ (of the background gauge superfield) and $q$ (of the quantum gauge superfield, ghosts, and matter superfields) should be set to 0 in order to get rid of the terms proportional to $\Lambda^{-n}$, where $n\ge 1$.

Certainly, from (\ref{NSVZ_Consequence}) one can easily obtain the NSVZ relation (\ref{NSVZ}) for the RG functions defined in terms of the bare coupling constants. It is important that the RG functions defined in terms of the bare charges are scheme-independent for a fixed regularization \cite{Kataev:2013eta}. That is why in the Abelian case Eq. (\ref{NSVZ}) is valid for an arbitrary choice of the renormalization prescription if the theory is regularized by the higher derivative method \cite{Kataev:2013eta,Kataev:2013csa,Kataev:2014gxa}. We believe that in the non-Abelian case Eq. (\ref{NSVZ_Consequence}) can be also derived using the BRST invariant version of the higher covariant derivative regularization. However, for RG functions defined in the standard way in terms of the renormalized coupling constants \cite{Bogolyubov:1980nc} it is necessary to specify the NSVZ subtraction scheme. This can be easily done starting from Eq.(\ref{NSVZ_Consequence}) repeating the argumentation of Ref. \cite{Kataev:2013eta,Kataev:2013csa,Kataev:2014gxa}. In the non-Abelian case the RG functions entering into Eq.(\ref{NSVZ_Consequence}) defined in terms of the bare coupling constant coincide with ones defined in terms of the renormalized coupling constants if the boundary conditions

\begin{equation}\label{NSVZ_Scheme1}
Z_\alpha(\alpha,\lambda,x_0) = 1;\qquad (Z_\phi)_i{}^j(\alpha,\lambda,x_0)=\delta_i{}^j;\qquad Z_c(\alpha,\lambda,x_0)=1,
\end{equation}

\noindent
where $x_0$ is a fixed value of $\ln\Lambda/\mu$, are imposed on the renormalization constants. (For example, it is possible and convenient to choose $x_0=0$.) Certainly, we also assume that the renormalization constants satisfy Eq. (\ref{Z_Constants_Natural}) and the equation

\begin{equation}\label{NSVZ_Scheme2}
Z_V = Z_\alpha^{1/2} Z_c^{-1},
\end{equation}

\noindent
which follows from the non-renormalization theorem derived in this paper. Thus, Eqs. (\ref{NSVZ_Scheme1}), (\ref{NSVZ_Scheme2}), and (\ref{Z_Constants_Natural}) presumably give the NSVZ scheme in the non-Abelian case if the supersymmetric gauge theory is regularized by the BRST invariant version of the higher covariant derivative regularization. This prescription is a straightforward generalization of the results of Ref. \cite{Kataev:2013eta,Kataev:2013csa,Kataev:2014gxa} for the Abelian case. It should be also noted that in the case of using the dimensional reduction so far there is no similar prescriptions which works in all orders \cite{Jack:1996vg,Jack:1996cn,Jack:1998uj,Harlander:2006xq,Mihaila:2013wma}, although the structures similar to integrals of $\delta$-singularities (which appear with the higher derivative regularization due to the  factorization of loop integrals into integrals of double total derivatives) were considered \cite{Aleshin:2015qqc}.

\section{Conclusion}
\hspace*{\parindent}

In this paper using the Slavnov--Taylor identities we prove that four three-point Green functions corresponding to the ghost-gauge vertices in ${\cal N}=1$ SYM theories with matter are finite. As a consequence, one can choose a subtraction scheme in which the corresponding renormalization constant is equal to 1. (Certainly, finite renormalizations are also possible.) In principle, this result can be considered as a non-renormalization theorem. Moreover, in this paper we argue that it could be useful for deriving the exact NSVZ $\beta$-function in the non-Abelian case. Really, qualitatively, in order to find the NSVZ relation we consider a graph without external lines. Then, attaching two external lines of the background gauge superfield one obtains diagrams contributing to the $\beta$-function. From the other hand, cutting the considered graph gives diagrams contributing to the anomalous dimensions of various fields, namely, the matter superfields, the quantum gauge superfield, and the Faddeev--Popov ghosts. The non-renormalization theorem proved in this paper allows writing the NSVZ $\beta$-function in the form of a relation between the $\beta$-function and the anomalous dimensions of these superfields. Thus, we obtain exactly the same qualitative picture for the origin of the NSVZ relation (for the RG functions defined in terms of the bare coupling constants) as in the Abelian case. If this picture is correct, the NSVZ scheme for the RG function defined in terms of the renormalized coupling constants can be constructed similar to the Abelian case, if the BRST invariant version of the higher covariant derivative method is used for the regularization.

\bigskip

\section*{Acknowledgments}
\hspace*{\parindent}

The author is very grateful to A.L.Kataev for valuable discussions. The work was supported by the RFBR grant No. 14-01-00695.

\appendix

\section{Appendix}
\hspace*{\parindent}

In this appendix we present one-loop expressions for the functions $f(p,q)$ and $F_\mu(p,q)$ defined by Eq. (\ref{Three_Point_Function1}).\footnote{All momentums in the equations (\ref{Function_f}) and (\ref{Function_F_mu}) are Euclidean.} Because they are rather large, here we will use the notation

\begin{equation}
\Delta_q \equiv \frac{\xi_0}{K_q} - \frac{1}{R_q}.
\end{equation}

\noindent
Then the results, which have been obtained by calculating the diagrams in Fig. \ref{Figure_VCC_Function}, can be written in following form:

\begin{eqnarray}\label{Function_f}
&& f(p,q) = \frac{1}{4}\int \frac{d^4k}{(2\pi)^4} \frac{e_0^2 C_2}{k^2 (k+q)^2 (k+q+p)^2} \Bigg\{\frac{2 k_\mu q_\mu}{(k+q)^2}\Delta_{k+q}
+ \frac{2k^2}{(k+q+p)^2} \Delta_{k+q+p} \nonumber\\
&& + R_p \Bigg(\frac{2k_\mu (q+p)^\mu}{(k+q+p)^2 R_{k+q}} \Delta_{k+q+p}
+ \frac{2k^2}{(k+q)^2 R_{k+q+p}} \Delta_{k+q}
+ \Big(\frac{k_\mu (k+q+p)^\mu}{(k+q+p)^2} \nonumber\\
&&  + \frac{k_\mu(k+q)^\mu}{(k+q)^2}\Big) \Delta_{k+q} \Delta_{k+q+p}\Bigg)
- \frac{2k_\mu (k+q)^\mu}{R_{k+q} R_{k+q+p}} \cdot \frac{R_{k+q+p}-R_{k+q}}{(k+q+p)^2-(k+q)^2}
\nonumber\\
&& - \frac{2(R_{k+q+p}-R_p)}{(k+q+p)^2-p^2}\cdot\frac{1}{R_{k+q+p}}\Bigg(\frac{k_\mu q^\mu (k+q+p)^2 - k_\mu q^\mu p^2}{(k+q)^2}
\Delta_{k+q} + \frac{k_\mu p^\mu}{R_{k+q}}\Bigg)
\nonumber\\
&& - \frac{2(R_{k+q}-R_p)}{(k+q)^2-p^2}\cdot\frac{1}{R_{k+q}}\Bigg(
\frac{k^2 (k+q)^2- k^2 p^2}{(k+q+p)^2} \Delta_{k+q+p}  + \frac{k_\mu (k+q)^\mu}{R_{k+q+p}}\Bigg)
\Bigg\} + O(e_0^4,e_0^2\lambda_0^2);\qquad
\end{eqnarray}

\begin{eqnarray}\label{Function_F_mu}
&& F_\mu(p,q) = \frac{1}{16}\int\frac{d^4k}{(2\pi)^4}\frac{e_0^2 C_2}{k^2 (k+q)^2 (k+q+p)^2}\Bigg\{
\frac{2}{k^2} \Delta_{k}\Big[ (q+p)_\mu\, k_\alpha (k+q)^\alpha
+ q_\mu\, k_\alpha
\nonumber\\
&& \times (k+q+p)^\alpha + k_\mu \Big(k^2 - q^2 - q_\alpha p^\alpha\Big)\Big]
- \frac{4 k_\mu}{R_{k+q}} + \frac{2}{(k+q)^2} \Delta_{k+q} \Big[ - q_\mu k_\alpha p^\alpha + p_\mu k^2
\nonumber\\
&& + k_\mu q_\alpha p^\alpha - k_\mu (k+q)^2 + k_\alpha q^\alpha (2q + 2k +p)_\mu\Big] + \frac{2}{(k+q+p)^2} \Delta_{k+q+p} \Big[
q_\mu k_\alpha (q+p)^\alpha\nonumber\\
&&  + (q+p)_\mu k_\alpha q^\alpha - k_\mu (q^2 + q_\alpha p^\alpha + k^2) - p_\mu k^2\Big] - \frac{R_{k+q+p}-R_{k+q}}{(k+q+p)^2-(k+q)^2}\cdot (2q + 2k + p)_\mu
\nonumber\\
&& \times \frac{4 k^\alpha q_\alpha}{R_{k+q} R_{k+q+p}}
+ \frac{2 R_p}{(k+q)^2 (k+q+p)^2} \Delta_{k+q+p} \Delta_{k+q} \Big[ (p_\mu p^\nu - \delta_\mu^\nu p^2) \Big((k^2 +q^2)(k_\nu + q_\nu)
\nonumber\\
&&  - (k+q)^2 q_\nu\Big) + p^2(q_\mu k_\alpha p^\alpha - k_\mu  q_\alpha p^\alpha)\Big]
+\frac{4 R_p}{(k+q)^2 R_{k+q+p}} \Delta_{k+q} \left(q_\mu k_\alpha p^\alpha - k_\mu q_\alpha p^\alpha\right)
\nonumber\\
&&  + \frac{4(R_{k+q}-R_p)}{(k+q)^2-p^2}
\frac{(k_\mu q_\alpha p^\alpha - q_\mu k_\alpha p^\alpha)}{R_{k+q} R_{k+q+p}}
+ \frac{4(R_{k+q+p}-R_p)}{(k+q+p)^2-p^2} \Bigg(\frac{(p_\mu p^\nu - \delta_\mu^\nu p^2) k_\nu}{R_{k+q+p} R_{k+q}} + \Delta_{k+q}\nonumber\\
&&\times \frac{\left((k+q+p)^2 -p^2\right)}{(k+q)^2 R_{k+q+p}}  \Big(q_\mu k_\alpha p^\alpha
- k_\mu q_\alpha p^\alpha \Big)\Bigg)\Bigg\}  + O(e_0^4,e_0^2\lambda_0^2).
\end{eqnarray}

\noindent
One can easily verify that these expressions are finite in the ultraviolet region, in agreement with the non-renormalization theorem proved in this paper. However, for these two functions in the one-loop approximation the result is trivial, because it can be derived from simple dimensional considerations.

\end{document}